\documentclass[12pt,a4paper]{article}
\usepackage[margin=1in]{geometry}
\usepackage{graphicx}        % standard LaTeX graphics tool when including figure files
\usepackage{multicol}        % used for the two-column index
\usepackage[bottom]{footmisc}% places footnotes at page bottom
\usepackage{amsfonts}
\usepackage[cmex10]{amsmath}
\usepackage{subcaption}
\usepackage{float}
\usepackage{bm}
\usepackage{setspace}
\usepackage{booktabs}
\usepackage{hyperref}

\usepackage{pgfplots}
\pgfplotsset{compat = newest}
\usepackage{tikz}
\usetikzlibrary{positioning,shapes.geometric,arrows.meta,3d, decorations.pathreplacing}

\definecolor{mypink}{HTML}{EE7993} % Pinkish-red
\definecolor{mygreen}{HTML}{3FB17D} % Green
\definecolor{myteal}{HTML}{79EEE2}
\definecolor{myyellow}{HTML}{F5D547} % Warm, sunny yellow
\usepackage{nicematrix}

\newcommand*\greencircle[1]{\tikz[anchor=mid]{% <---- BEWARE
\node[shape=circle,draw,inner sep=2.5pt,fill=mygreen,yshift=1em] (char) {\text{\footnotesize #1}};}}
\newcommand*\pinkcircle[1]{\tikz[anchor=mid]{% <---- BEWARE
\node[shape=circle,draw,inner sep=2.5pt,fill=mypink,yshift=1em] (char) {\text{\footnotesize #1}};}}
\newcommand*\blankcircle[1]{\tikz[baseline=(char.base)]{% <---- BEWARE
\node[shape=circle,draw,inner sep=1.5pt,fill=white,yshift=1em] (char) {\text{\footnotesize #1}};}}

\newcommand*\yellowcircle{\tikz[anchor=mid]{% <---- BEWARE
\node[shape=circle,draw,inner sep=2.5pt,fill=myyellow,yshift=1em] (char) {~};}}

\newcommand*\tealcircle{\tikz[anchor=mid]{% <---- BEWARE
\node[shape=circle,draw,inner sep=2.5pt,fill=myteal,yshift=1em] (char) {~};}}

\usepackage{nicematrix}
\usepackage{tikzpeople}
\usepackage{fontawesome5}

\DeclareRobustCommand*\alice{\tikz[scale=0.8]{\node[alice] at (0,0) {};}}
\DeclareRobustCommand*\bob{\tikz[scale=0.8]{\node[bob] at (0,0) {};}}

\newcommand{\customfootnotetext}[2]{{% Group to localize change to footnote
  \renewcommand{\thefootnote}{#1}% Update footnote counter representation
  \footnotetext[0]{#2}}}% Print footnote text

\usepackage[backend=biber, style=authoryear, sortcites=false,sorting=none,uniquelist=false,natbib=true,doi=true,isbn=false,url=false,eprint=false,pagetracker, ibidtracker=constrict, giveninits=true,maxbibnames=10,maxcitenames=1,date=year]{biblatex}
\AtEveryBibitem{\clearfield{month}}

\begin{document}

%TC:ignore

% \title{Inferring conventions' propagation mechanisms from behavioral network data}
% \title{``When her family finds [out] you are using the wrong metric\dots'': dilemmas and trade-offs in the diffusion of conventions}
\title{Dilemmas and trade-offs in the diffusion of conventions}
\date{}
% \title{}

%
%                                     also used for the TOC unless
%                                     \toctitle is used
%
\author{Lucas Gautheron\textsuperscript{1,2}\\
\texttt{lucas.gautheron@gmail.com}%\\
%\texttt{lucasgautheron.github.io}
}

\maketitle

% \paragraph{Acknowledgements} Many thanks to Radin Dardashti, Thomas Heinze, Olivier Morin, and Cailin O'Connor for their feedback. I am also grateful to Jeffrey Barrett, Sean Carroll, and Brian Skyrms for inspiring discussions. Finally, I would like to thank Robert D. Hawkins for further suggestions.
% % \paragraph{Acknowledgements} [ANONYMIZED]

% \paragraph{Code and data} The code and data for this paper is available at \url{https://gin.g-node.org/lucasgautheron/dilemmas-conventions}.
% % \paragraph{Code and data} The code and data for this paper is available at [ANONYMIZED].

% \paragraph{Funding} The author acknowledges funding from the DFG Research Training Group 2696 ``Transformations of Science and Technology since 1800''.
% % \paragraph{Funding} [ANONYMIZED]

% \paragraph{Competing interests} The author declares a personal inclination towards the $(+,-,-,-)$ metric signature.
\customfootnotetext{1}{Evolution, Science and Society, University of Missouri, Columbia, United States}

\customfootnotetext{2}{Interdisciplinary Centre for Science and Technology Studies (IZWT), University of Wuppertal, Germany}

% \customfootnotetext{3}{Département d'Études Cognitives, École Normale Supérieure, Paris, France}

 %
% \textbf{Word count}: \quickwordcount{main}
%TC:endignore

% \newpage

% {\Large{Dilemmas and trade-offs in the diffusion of conventions}}

\begin{abstract}
Outside ideal settings, conventions are shaped by heterogeneous competing processes that can challenge the emergence of norms. In order to acknowledge this complexity, this paper develops a generalized account of conventions and identifies three trade-offs involved in their diffusion: (I) the trade-off  between the imperatives of social, sequential, and contextual consistency that individuals balance when choosing between conventions; (II) the competition between local (bottom-up) and global (top-down) coordination, depending on whether individuals coordinate their behavior via interactions throughout a social network or external factors transcending the network; and (III) the balance between decision optimality (e.g., collective satisfaction) and decision costs when collectives with conflicting preferences choose a convention. A broadly applicable statistical physics framework for exploring these trade-offs is developed and applied to a sign convention in physics. The method can infer the structure of the underlying coordination game, the networks of social interactions involved, and the processes through which conflicts are resolved. This shows that the purpose of conventions may exceed coordination, and that individual preferences towards conventions are concurrently shaped by cultural factors and multiple social networks. Finally, this work emphasizes the role of leadership in the resolution of conflicts. 
\end{abstract}


\section{Introduction}

Since David Lewis \citep{Lewis2002}, conventions (including linguistic norms, technological or manufacturing standards, and many other social norms) are primarily conceived as solutions to coordination problems \citep{hawkins2019emergence}. Yet, the attitude of individuals towards conventions involves a multitude of factors beyond social coordination, resulting in tensions that may disrupt the emergence of a universal norm. To acknowledge this complexity, this paper develops a generalized statistical physics account of conventions and identifies three trade-offs involved in their diffusion and the resolution of conflicts in the absence of consensus. The first trade-off is the balance between i) social consistency (driven by coordination with peers), ii) sequential consistency (driven by the cost of switching from one practice to another), and iii) contextual consistency (driven by a need for consistency with other choices or cultural traits) (\S\ref{section:social_sequential_contextual}). The second trade-off involves the balance between \textit{local} (or bottom-up) versus \textit{global} (or top-down) coordination, depending on whether individual preferences are formed endogenously through local interactions on a social network, or by factors transcending the network structure (or both, in possibly contradicting ways) (\S\ref{section:local_global}). Finally, the last trade-off is the balance between decision costs and the optimality of the outcome in the resolution of conflicts (\S\ref{section:optimality_costs}). To illustrate these trade-offs, we apply a statistical physics framework to behavioral data about a sign convention in physics. %This convention involves a choice between two physically equivalent options for describing the metric properties of space-time: the ``mostly plus'' ($+1$) and the ``mostly minus'' ($-1$) metric signatures. %We rely on authorship, citation, and textual data extracted from Inspire HEP, a comprehensive database of high-energy physics literature. 
This statistical framework allows us to retrieve information about individuals' decision-making, the structure of the underlying coordination problem, and the multiple infrastructures (whether social or cultural) involved in the propagation of a convention.

First, we show that scientists' attitude is shaped by sequential consistency, as they tend to maintain a preferred choice in their solo-authored publications independently of the target research area (\S\ref{section:consistency_context}). Then, we show that scientists' preferences are correlated -- albeit imperfectly -- with those of their co-authors, which means that some level of social coordination is achieved (\S\ref{section:local_versus_global}). In order to explain how, the relative contribution of local coordination (via dyadic interactions with peers) and global coordination (i.e. via shared culture) is measured by applying the Ising model to the authors' collaboration and citation network. This shows that both local and global processes contribute to coordinating scientists' preferences. %The Ising model approach can recover the structure of the underlying coordination games, as well as the relative contribution of multiple social networks to the emergence of coordination. 
Third, we assess the plausibility of three mechanisms of preference-formation according to their ability to explain the observed magnitudes of local and global coordination, and find slightly more evidence for a model of cultural transmission involving the imitation of peers (\S\ref{section:inferring}). Finally, we infer the process through which scientists resolve conflicts about which convention to use in collaborations (\S\ref{section:conflict_resolution}). We find evidence that the last author's preference most often prevails, thus highlighting the role of seniority and power in the resolution of conflicts. Taken together, these results indicate that decision-making processes related to conventions involve multiple and sometimes conflicting factors.  %This requires a generalization of Lewis' coordination-based account.
%This requires an expanded, multi-dimensional account of conventions, beyond the focus on coordination alone.

\subsection{Background}

In relativistic physical theories (such as general relativity and quantum field theory), the ``metric tensor'' is a mathematical object that represents the metric properties of space-time. %It can be seen as a matrix that defines a pseudo-distance between events according to their time and space coordinates. In particular, in the vacuum, one can choose the metric tensor to take either of the following forms:
Broadly speaking, the metric tensor can take either of the two following forms:

\begin{equation*}
\begin{pmatrix}
+1 & 0 & 0 & 0\\
0 & -1 & 0 & 0\\
0 & 0 & -1 & 0\\
0 & 0 & 0 & -1\\
\end{pmatrix} \text{ or } \begin{pmatrix}
-1 & 0 & 0 & 0\\
0 & +1 & 0 & 0\\
0 & 0 & +1 & 0\\
0 & 0 & 0 & +1\\
\end{pmatrix}
\end{equation*}

The first choice $(+,-,-,-)$ is known as the mostly minus convention while the second choice, $(-,+,+,+)$ is referred to as the mostly plus convention. These choices are regarded as physically equivalent and lead to identical predictions. However, depending on which convention is chosen, certain quantities arising in calculations will take either positive or negative values. Interestingly, there is no norm and both conventions are used. This creates a puzzle: if, as per Lewis, scientific conventions address an imperative of social coordination, why physicists persist in making opposite choices?
%As we shall see, our framework can leverage this diversity of behavior to infer the mechanisms driving physicists' attitudes towards this convention.

The puzzle points to a gap in the existing theoretical toolkit: while formal models have provided rich insights into norms and conventions, they may have left out crucial aspects of reality by stripping away too much of its complexity \citep{Elsenbroich2013}, or by neglecting the interactions between phenomena studied in isolation\footnote{For instance, \citep{Delgado2002,pujol2005role} demonstrated the sensitivity of conventions to the topological features of the underlying social networks (including their small-world, scale-free or clustering properties).}. Similarly, while controlled experiments can uncover certain aspects of conventions in idealized settings \citep{Guala2010,Bruner2014,Centola2015,Hawkins2016,Formaux2021,hawkins2023partners}, they may conceal the fact that complex heterogeneous processes and multiple social infrastructures can drive or prevent the emergence of conventions in naturalistic situations \citep{Boyce2024}. Fortunately, the advent of large online communities has opened up opportunities to investigate the diffusion of real norms and conventions in complex networks \citep{DanescuNiculescuMizil2013,kooti2012emergence,Heaberlin2016,Rotabi2017}. Such data-driven approaches, however, have barely extended to the study of scientific conventions\footnote{with the exception of \citep{Rotabi2017}, who investigated LaTeX macros naming conventions in scientific papers.}. Yet, ``conventionalism'' can be traced back to Poincaré and his account of the epistemic status of the axioms of geometry \citep{ben2006conventionalism,Wilholt2012-WILCPD}. In fact, conventions are ubiquitous in science \citep{Wu2023}, including statistical practices such as statistical significance thresholds \citep{wilholt2009bias}, measurement strategies \citep{Smaldino2022}, and unit systems. %, scientific jargon, and any other conventional means of asserting reproducibility and mutual understanding.
%Nevertheless, previous culturally evolutionary perspectives on science \citep{Wu2023} have not connected formal models of conventions to empirical data from scientific settings, a gap that this paper addresses.

By exploring a scientific convention -- namely, the metric signature --, this work highlights the interactions between multiple phenomena involved in the diffusion of conventions that prior works have addressed separately or ignored, and provides cues for understanding how conventions can fail to develop into universal norms, in naturalistic settings. This task, as we will now show, calls for an expanded account of conventions.

%First, we identify three hypothetical trade-offs affecting conventions in general and investigate each of these empirically using a sign convention in physics. To this end, we propose a broadly applicable statistical physics framework, such that our approach can capture the postulated trade-offs in a large array of contexts. Therefore, while the convention studied in the present paper is rather innocuous, this work paves the way for further empirical investigation of norms and conventions, in science or beyond. 
%For instance, coordination among peers (which should play a major role according to most accounts of conventions) matter less to scientists' behavior than might have been expected from the traditional account. On the other hand, scientists' behavior suggests that they strongly value \textit{internal} consistency, i.e., the need to appeal to the same convention throughout their works.

\subsection{\label{section:social_sequential_contextual}The trade-off between social, sequential, and contextual consistency}

While Lewis' account of conventions is focused on their social dimension, earlier accounts provide different perspectives: holist approaches to conventionalism, for instance, emphasize that scientists have ``discretion'' to choose between distinct but collectively coherent systems of beliefs \citep[``Conventionalism as the Underdetermination of Theory'', pp.~8--12]{ben2006conventionalism}. Below, these two perspectives on conventions are brought together under a broader notion of collective consistency, a mereological description formalized using elements of game theory and statistical physics. This reveals that conventions involve multiple dimensions that can compete with each other. The statistical nature of the framework will enable us to infer the drivers underlying the adoption of a convention from behavioral patterns.

% Below, we elaborate an utilitarian description of the effect of social, sequential, and contextual consistency on individuals' decision-making. By analogy with statistical physics, we translate this description into probabilistic models suitable for empirical exploration. For simplicity, we assume individuals must choose between two options ($x=\pm 1$).

% \linespread{1.75}\selectfont
\begin{table*}[]
\centering
\begin{subtable}{0.3\textwidth}
\centering
\resizebox{!}{0.75cm}{\begin{NiceTabular}{c|c|c}[hvlines]
\diagbox{\tikz{\node[alice] at (0.5,0) {};}}{\tikz{\node[bob] at (-0.5,0) {};}} & $x_{j}=\pinkcircle{~}$ & $x_{j}=\greencircle{~}$ \\
$x_{i}=\pinkcircle{~}$ & $(1,1)$ & $(0,0)$ \\
$x_{i}=\greencircle{~}$ & $(0,0)$ & $(1,1)$ \\
\end{NiceTabular}}
\caption{\protect\label{table:social}\textbf{Social consistency}. Alice and Bob are better off if they agree on either \pinkcircle{~} or \greencircle{~}.}
\end{subtable}
\hfill
\begin{subtable}{0.3\textwidth}
\centering
\resizebox{!}{0.75cm}{\begin{NiceTabular}{c|c|c}[hvlines]
\diagbox{\tikz{\node[alice] at (0.5,0) {};}}{\tikz{\node[alice] at (-0.5,0) {};}} & $x_{t+1}=\pinkcircle{~}$ & $x_{t+1}=\greencircle{~}$ \\
$x_{t}=\pinkcircle{~}$ & $1$ & $0$ \\
$x_{t}=\greencircle{~}$ & $0$ & $1$ \\
\end{NiceTabular}}
\caption{\protect\label{table:sequential}\textbf{Sequential consistency}. Alice is better off if she consistently chooses \pinkcircle{~} or \greencircle{~}.}
\end{subtable}
\hfill
\begin{subtable}{0.3\textwidth}
\centering
\resizebox{!}{0.75cm}{\begin{NiceTabular}{c|c|c}[hvlines]
\diagbox{\tikz{\node[alice] at (0.5,0) {};}}{\tikz{\node[alice] at (-0.5,0) {};}} & $y=\yellowcircle{}$ & $y=\tealcircle{}$ \\
$x=\pinkcircle{}$ & $1$ & $0$ \\
$x=\greencircle{}$ & $0$ & $1$ \\
\end{NiceTabular}}
\caption{\protect\label{table:contextual}\textbf{Contextual consistency}. Alice is better of if she chooses either \pinkcircle{}\yellowcircle{} or \greencircle{}\tealcircle{}.}
\end{subtable}
\caption{\label{table:consistency}Collective consistency as coordination games involving Alice (\alice) and Bob (\bob), or Alice alone. Each table represents a payoff matrix associated with a ``collective'' choice.}
\end{table*}

\paragraph{\label{paragraph:coordination}Social consistency and coordination costs} Conventions are mainly conceived as solutions to coordination problems \citep{Lewis2002}, which arise when individuals would benefit from acting in a mutually consistent way, but struggle to do so -- maybe, for instance, because they lack the information necessary for achieving joint-action \citep{Lewis2002,hawkins2019emergence}. Conventions can solve coordination problems by providing individuals with expectations about how others will behave in a given setting, a paradigmatic example being left-hand versus right-hand traffic. In absence of universal conventions, individuals experience \textit{coordination costs} in their interactions. When interactions involve two people at a time, coordination costs can be represented by a payoff matrix that defines the utility (i.e. the rewards) $u_{i,j}(x_i,x_j)$ for agents $i$ and $j$ as a function of $x_i$ and $x_j$, their respective strategies (see Table \ref{table:social}; for clarity, we consider binary conventions, labeled by $x\in\{-1,+1\}$). %
Additionally, coordination costs are specified by a network structure capturing the frequency of interactions $w_{ij}$ between any pair $(i,j)$ of agents. To see how such a system may spontaneously evolve, one can 
construct a ``potential'' $U(x_1,\dots,x_N)$, i.e. a function of the joint strategy of every individual $1\leq i \leq N$ \citep{Szab2016}. The potential is established by constructing a  function that varies by an amount $\Delta(x_i\to x_{i'})=\sum_j w_{ij} [u_i(x_i',x_j)-u_i(x_i,x_j)]$ as an agent $i$ unilaterally changes their strategy from $x_i$ to $x_i'$ \citep{Szab2016}. Under a class of noisy best-response strategies, where noise is a proxy for bounded rationality\footnote{For ``potential'' games, the ``logit'' rule and the Glauber dynamics lead to the above Boltzmann distribution \citep{Szab2016,Perc2017}. These rules correspond to a certain class of strategies, where the players do not systematically choose the best response but instead choose random responses with probabilities that are increasing functions of their expected utility \citep{Rosenthal1989,McKelvey1995}.}, it can be shown that the probability of a particular combination of individual strategies becomes:

\begin{equation}
    \label{eq:coordination}
    P(x_1,\dots,x_N)=\dfrac{1}{Z}e^{\beta U(x_1,\dots,x_N)}
\end{equation}

Where $Z$ is a normalization constant and $\beta \geq 0$ controls the degree of rationality -- or efficiency -- of the agents \citep{Correia2022,Zimmaro2024}. In statistical physics, \eqref{eq:coordination} is the Boltzmann distribution; $U$ is (up to a minus sign) the energy potential of a particular configuration, and $\beta$ is the inverse temperature\footnote{Often, $\beta$ may be omitted without loss of generality through proper rescaling of $U$.}. In the case of symmetric coordination games (Table \ref{table:social}), this gives:
\begin{equation}
    P(x_1,\dots,x_N)=\dfrac{1}{Z}e^{\frac{\beta}{2} \sum_{ij} w_{ij} x_i x_j}
\end{equation}

%Interestingly, this probabilistic framework enables the retrieval of information about the coefficients of the payoff matrices ($u_{i},u_{j}$) or the network structure ($w_{ij}$) from observations of individuals' strategies, as shown in \S\ref{section:local_versus_global}. 

This distribution is the equilibrium configuration of the Ising model \citep{Macy2024}: it describes physical systems composed of many interacting magnetic ``spins'', that are more stable when neighboring spins are aligned in the same direction, regardless of which particular orientation. 

The parallel is both insightful and practical. First, the statistical model enables empirical explorations of conventions: given observations of individual strategies, it can be used to retrieve the structure of the underlying game or identify the relevant social interaction network(s). Second, as we will see, the Ising model possesses extensions capable of capturing additional influences over conventions, such as factors external to social interaction networks.
 %Additionally, the degree of social consistency achieved for such binary conventions can be measured via $\langle x_i x_j \rangle = \sum_{i,j} w_{ij} x_i x_j/\sum_{i,j} w_{ij}$, a quantity comprised between -1 (perfect anti-alignment) and +1 (perfect alignment).

\paragraph{\label{paragraph:consistency}Sequential consistency and switching costs}

In addition to addressing coordination problems, conventions enable individuals to settle on one choice once and for all, in a way that facilitates future moves. Consider keyboard layouts (e.g. qwerty). While there exists many such layouts, we benefit from settling on a single one, even if our choice is arbitrary and distinct from our peers'.  In that respect, certain conventions can serve a purely internal purpose of consistency, as if individuals ``played'' a coordination game with themselves, such that their payoffs depend solely on the mutual coherence of their own actions.  In science for instance, a convention (such as a notational choice) may be established locally, in the context of a single paper, merely to establish consistency between different calculations. Sequential consistency can thus be orthogonal to, or even antagonistic with the need to conform to a social environment.

To model sequential consistency, let $x_{it}$ be the convention employed by agent $i$ at time $t\in\{1,\dots,T\}$. A simple model of the utility of a sequence of choices for an isolated individual is $U(x_{i1},\dots,x_{iT})=\sum_{t=1}^{T-1} u(x_{i,t},x_{i,t+1})$, where $u(x_t,x_{t+1})$ is the payoff matrix associated with the transition from $x_t$ to $x_{t+1}$ (Table \ref{table:sequential}). In such a model, agents experience costs every time they switch from one convention to another. We may again assume thats agents behave semi-rationally, such that $p(x_{t+1}|x_t)=e^{\beta x_{t}x_{t+1}}/(e^{-\beta x_{t}x_{t+1}}+e^{\beta x_{t}x_{t+1}})$, with $\beta>0$ their degree of rationality. We obtain a Markov model with transition matrix $p(x_{t+1}|x_t)$.
%This model alone can explain path-dependency, if agents are reluctant to deviate from their first move.
Alternatively, sequential consistency may reflect lasting preferences with memory effects due to complex long-range interactions between individual actions. Without loss of generality, we can express 
 the probability of a particular sequence as $P(x_{i1},\dots,x_{iT})\propto e^{\beta U(x_{i1},\dots,x_{iT})}$, where $\beta$ is, as before, a measure of efficiency, and $U$ is utility of a sequence of decisions. This suggests a class of model for studying the evidence for sequential consistency in behavioral patterns. %For practical reasons, the latent utility $U$ can be assumed to take the form $U(x_{i1},\dots,x_{iT})=\sum_{t=1}^T u_i^{x_{it}}$ where $u_i^{x}$ designates the utility associated with choice $x$ for agent $i$. 
 %Similarly, the degree of sequential consistency can be crudely estimated via $\langle x_{i,t} x_{i,t+1}\rangle = \sum_i \sum_t x_{i,t} x_{i,t+1}/\sum_{i}\sum_t \bf{1}$.

\paragraph{\label{paragraph:adaptation}Contextual consistency and maladaptation costs} Some conventions are less conventional than others \citep{OConnor2020,Gasparri2023}: certain choices can be \textit{maladaptive} and less likely to be adopted. However, which conventions are more or less adaptive may depend on the context. Unit systems are a good example: while light-years might be a convenient unit of length for astronomers, engineers may reasonably prefer meters. Maladaptation costs indicate an inconsistency between a convention and other interacting choices or cultural traits. This can be thought of in terms of a cultural \citep{Poulsen2023} or epistemic \citep{alexander2015epistemic} fitness landscape, where $U(x_{i,1}, \dots, x_{i,C})$ describes the fitness of a configuration of $C$ traits $\bm{x}_i=(x_{i,1},\dots,x_{i,C})\in\{\pm 1\}^C$. It is possible that the choice between, say, $x_{i,1}=-1$ or $+1$ is ``conventional'', in that there is no universally superior choice across the landscape (i.e. $\mathbb{E}_{\bm{x}\sim p(\bm{x}|x_1=+1}[U(\bm{x})|x_1=-1] \simeq \mathbb{E}_{\bm{x}\sim p}[U(\bm{x})|x_1=+1]$, where $p$ is the joint probability distribution over all traits), even though certain regions in the landscape may locally favor a specific choice for $x_1$\footnote{This is obvious in the context of language. The mapping between objects and symbols is highly conventional; however, for a given pre-existing language, the choice of how to name a new object can be constrained or suggested by preceding linguistic infrastructure.}. The cultural fitness landscape can be modeled using the same building blocks as for social and sequential consistency, given pairwise interactions between cultural traits encoding the need for mutual consistency (Table \ref{table:contextual}). In fitness landscapes, such interactions are generally referred to as ``epistasis'' \citep{Pascual2020}. As \citealt{alexander2015epistemic} put it, ``one can think of the amount of epistasis in a fitness function [\dots] as a formal model of the Quinean web of belief''. Therefore, contextual consistency captures the holistic dimension of conventions: that is, the freedom to choose between systems of beliefs, as long as those beliefs are internally coherent.

Under the assumption that traits are pairwise-interacting, and further assuming that they co-evolved according a certain class of evolutionary rules\footnote{e.g. Glauber dynamics, cf. \citealt{Walter2015}}, their distribution converges to the Ising model equilibrium. Therefore, cultural fitness landscapes involving multiple traits and conventions can sometimes be reconstructed empirically from behavioral patterns by solving an inverse Ising problem  \citep{Poulsen2023}. In certain cases, this can reveal a plurality of collectively consistent systems of choices: a demonstration is proposed in S\ref{landscapes}, using collections of naming conventions in a scientific typesetting language. Under this approach, mutually consistent systems of decisions visually materialize as modes (or peaks) in the empirically reconstructed fitness landscape. %When the position of agent $i$ in the landscape can be considered fixed, except for a focal trait $k$, the relative reward for their choice $x_{i,k} \in\{\pm 1\}$ reduces to $U(x_{i,1},..,x_{i,k},..,x_{i,C})-U(x_{i,1},..,-x_{i,k},..,x_{i,C}) \equiv B_i x^i_k = \pm B_i$. 
%The cultural landscape can alternatively be thought as a web of beliefs, in which whether certain beliefs are true is strongly contingent on the entire system of beliefs one adopts \citep{quine1978web}.
% Depending on their position in the ``cultural landscape'', an agent $i$ may experience different payoffs $B_i x_i=\pm B_i$ for each convention. Agents in the same region of the landscape may converge in behavior for reasons independent of social coordination (which corresponds to $B_i=B_j$ and $J=0$ in Table \ref{table:coordination_game}). In this case:

% \begin{equation}
%     \label{eq:solo}
%     P(x_i=x) = \frac{e^{\beta B_ix}}{e^{\beta B_ix}+e^{-\beta B_ix}}
% \end{equation}

\paragraph{Conventions as collective consistency: a mereological account}

Broadly speaking, conventionality arises when behavior is determined by ``collective'' rather than individual constraints: in a statistical sense, the marginal probability of a particular outcome $p(x_i)$ is weakly constrained (i.e., it has large entropy; \citealt{OConnor2020}); only the joint probability of all outcomes $p(x_1,\dots,x_n)$ is constrained, due to synergistic interactions between individuals, cultural traits, or consecutive choices. For example, in the case of sequential consistency, the first move does not matter, as long as the \textit{entire} sequence of actions is collectively consistent.  Contextual consistency is also a collective constraint, since it cannot reject a particular choice independently from other interacting choices. In any case, this property can be modeled using the same fundamental game-theoretic building blocks (Table \ref{table:consistency}) for encoding interactions between individual choices. The proposed framework is thus a mereological generalization of Lewis' coordination-game account of conventions: it proposes that conventionality reflects a certain kind of interdependency between individual choices resulting in a constraint on their joint distribution, regardless of whether said interdependencies reflect social coordination, sequential decision-making, or interacting traits. %The central assumption is that behavioral patterns reflect two components: a latent utility $U(\dots)$, encoding social, sequential, and contextual interactions between choices; and a noise parameter, $\beta$, which models bounded rationality.

%While this table features two-action two-person games, we may $n-$person games and more complex structures (hypergraphs rather than graphs) in order to fully characterize collective consistency.
% (for instance, epistemological holism states that the truth-value of certain statements cannot be determined independently from the truth-values of a web of other interconnected statements; in other words, in depends on the system of beliefs one adopts \citep{quine1978web})
%To put it briefly, coordination, sequential consistency, and maladaptation are respectively social, sequential, and contextual constraints on behavior, all of which are collective rather than individual.

\paragraph{Inferring the imperatives driving a particular convention}

In the most general case, all three factors (social, sequential, and contextual consistency) may be involved in conventions, albeit to varying extents. For the metric signature, coordination costs are plausible (it should be easier to collaborate with scientists who will systematically agree to using your favorite convention, and it is easier to copy results if they are systematically derived with the same convention). Switching costs are seemingly plausible, as working with different metric signatures implies keeping track of which sign certain quantities must take according to which convention is used. %In fact, publications generally use a single metric signature, which in itself is evidence for the value of sequential consistency (nothing in principle prevents different choices from being made in the same published unit). 
Finally, context-dependent maladaptation costs might be involved too. For instance, for problems that involve ``proper time'' calculations, the mostly minus metric is advantageous, since then proper time is equal to the pseudo-distance between events rather than minus the pseudo-distance. For any particular convention, which of these three imperatives prevail is an empirical question.

How can we infer which of these matters from behavioral patterns? A model incorporating all three factors and inspired from opinion-dynamics literature is confronted to the data in \S\ref{section:inferring} (see \S\ref{section:strategic_agent_model} for more details about the model itself). We can however simplify the model in numerous ways. For instance, we can leverage the fact that, when the position of agent $i$ in the underlying cultural landscape can be considered fixed, except for a focal trait $k$, the relative reward for their choice $x_{i,k} \in\{\pm 1\}$ reduces to $U(x_{i,1},..,x_{i,k},..,x_{i,C})-U(x_{i,1},..,-x_{i,k},..,x_{i,C}) \equiv B_i x^i_k = \pm B_i$. Thus, we can conveniently parameterize the effect of one choice over the other by a single scalar, $B_i$, encoding the effect of the context in which the agent operates their decision. We will repeatedly use this simplifying assumption by considering that the research area encodes all the contextual information relevant to the choice of a metric signature. Under this assumption, the effect of the cultural context can be modeled simply by updating the structure of the social coordination game from Table \ref{table:social}. This gives the payoff structure in Table \ref{table:coordination_game}, where the benefits of social coordination are parameterized by $J$, and the effects of the context is parameterized by $B$. When different players must navigate different contexts that suggest conflicting choices (i.e. when $B_i$ and $B_j$ have opposite signs), the game becomes asymmetric and possesses a ``battle of the sexes'' structure, capturing the tension between social consistency and contextual consistency. 

\begin{table}[!htb]
    \centering
     \caption{Generic payoff matrix of a two-player two-action coordination game. Cells indicate $(u_i(x_i,x_j), u_j(x_i,x_j))$, the rewards of $i$ and $j$ as a function of their joint strategy. $J>0$ measures the synergistic benefit of coordination, and $(B_i,B_j)$ measures the ``preferences'' of $i$ and $j$, due to (for instance) their positions in the cultural landscape.}
  
        \begin{tabular}{c|c|c|}
        \cline{2-3}
                                  & $x_j=\pinkcircle{~}$ & $x_j=\greencircle{~}$ \\ \hline
        \multicolumn{1}{|c|}{$x_i=\pinkcircle{~}$} & $(+J-B_i,+J-B_j )$   & $(-J-B_i,-J+B_j)$   \\ \hline
        \multicolumn{1}{|c|}{$x_i=\greencircle{~}$} & $(-J+B_i,-J-B_j)$   & $(+J+B_i,+J+B_j)$   \\ \hline
        \end{tabular}
        \label{table:coordination_game}
\end{table}

Under such a game structure, the joint probability over individual strategies becomes a simple variant of the Ising model \citep{Zimmaro2024}: 

\begin{equation}
    \label{eq:ising_asymmetric}
    P(x_1, \dots, x_N) = \dfrac{1}{Z}e^{\frac{\beta}{2} J\sum_{i,j} w_{ij}x_i x_j + \sum_i k_i B_i x_i}
\end{equation}

% In \S\ref{section:consistency_context}, we start by evaluating the importance of sequential and contextual consistency in the case of the metric signature. It will be shown that both matter, but sequential consistency matters more, such that individuals tend to stick to their favorite convention even as they publish across different contexts. Physicists therefore have \textit{preferences} towards a metric signature, and we may then ask how these preferences are formed. In \S \ref{eq:ising}, we also examine the relative contribution of social and contextual consistency in the formation of scientists' preferences.  

Where $k_i=\sum_{j}w_{ij}$ is the degree of $i$ in the network. Interestingly, parameters $J$ and $B$ from Table \ref{table:coordination_game} and eq. \eqref{eq:ising_asymmetric}, which follow directly from our account, reveal an interesting competition between what \textit{local coordination}, carried out by the social network $(w_{ij})$, and \textit{global} coordination, carried by the external contribution $B$.

%In the mostly minus convention, the pseudo-norm of particles' momentum four-vector is equal to the square of their mass (as opposed to minus their mass squared), which may be convenient in phenomenological calculations.
%Conversely, in dimensions greater than $3+1$, the sign of the determinant of the metric tensor changes depending on the number of dimensions with the mostly minus signature, and the mostly plus signature. None of these issues are insurmountable, but they introduce small context-dependent costs specific to each convention.

%When all individuals behave consistently and uniformly, it is difficult to infer which factor was determinant in the adoption of a norm. However, when variations are observed, they can be leveraged by statistical models to infer which of the above imperatives can better account for attitudes towards a convention.

\subsection{\label{section:local_global}Local and global processes in the diffusion of conventions}

The emergence of social norms is the byproduct of both ``local'', ``dyadic'' processes and pre-existing ``broader population-level infrastructure'' \citep{hawkins2019emergence,young1996economics}. In particular, we propose a distinction between \textit{local} and \textit{global} processes of coordination. ``Local'' coordination refers to bottom-up coordination via local interactions on a network (e.g. by the imitation of peers \citep{Moore2012}, or strategic adjustment to their behavior), as opposed to ``global'', top-down processes resulting from external factors transcending the network structure, including institutions, ``central authorities'' \citep{young1996economics}, but also any pre-established cultural traits or common knowledge shared within different groups. Global coordination can arise when individuals share the understanding that one option is intrinsically superior ($\mathrm{sign}(B_i)=\mathrm{sign}(B_j)$ and $|B_i|,|B_j\gg 1$ in Table \ref{table:coordination_game}). In scientific communities, local processes may propagate over a co-authorship network, while global factors may include a shared ``disciplinary matrix'' \citep{kuhn1970} sustained by various means including disciplinary institutions or textbooks
\footnote{This distinction between local and global differs from that suggested in \citep{Rotabi2017}, which opposes dynamics in the diffusion of conventions at the microscopic and macroscopic (or aggregate) levels. }. Disciplinary matrices are the basis for common knowledge in a research area; they constitute what every individual part of a research endeavor knows and expect others to know as well. Such higher-order type of knowledge is central to conventions \citep{Lewis2002}. Salience is another device susceptible of promoting global coordination. \citep{Lewis2002}.
%The local/global distinction resembles the endogeneous/exogeneous distinction made in previous works exploring the dynamics of collective attention in social media (e.g. \citep{lehmann2012dynamical}), where ``endogenous'' refers to behavior driven by interactions on a social network, and ``exogenous'' refers to processes dictated by external factors such as the mass media.

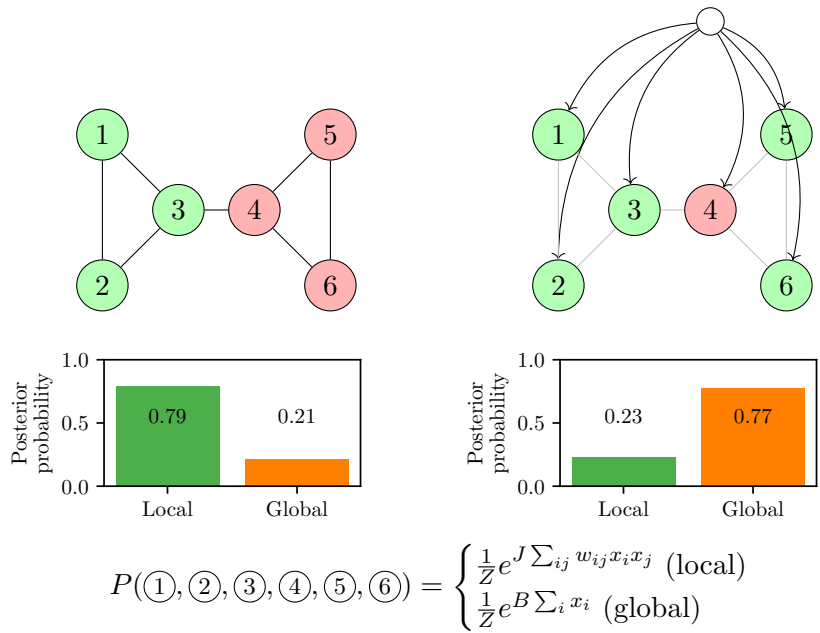
\begin{figure}[!h]
    \centering
\begin{tikzpicture}
    \node[circle, draw, fill=red!30] (D) at (2,2) {5};
    \node[circle, draw, fill=red!30] (E) at (2,0) {6};
    \node[circle, draw, fill=red!30] (F) at (1,1) {4};

    % Left graph (traditional)
    \node[circle, draw, fill=green!30] (A) at (-1,2) {1};
    \node[circle, draw, fill=green!30] (B) at (-1,0) {2};
    \node[circle, draw, fill=green!30] (C) at (0,1) {3};

    \draw (A) -- (B);
    \draw (B) -- (C);
    \draw (C) -- (A);
    \draw (D) -- (E);
    \draw (E) -- (F);
    \draw (F) -- (D);

    \draw (F) -- (C);

    \node at (0.1, -2) {\includegraphics[width=5cm]{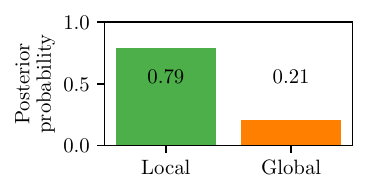}};
    % \node at (0.5, -4) {$P(\greencircle{1},\greencircle{2},\greencircle{3},\pinkcircle{4},\pinkcircle{5},\pinkcircle{6})=\begin{cases}\frac{1}{Z}e^{-J\sum_{ij} w_{ij} x_ix_j}\\\frac{1}{Z}e^{-B\sum_{i}x_i}
    %  \end{cases}$};

    % Right graph (common ancestor)
    \node[circle, draw, fill=green!30] (D1) at (8,2) {5};
    \node[circle, draw, fill=green!30] (E1) at (8,0) {6};
    \node[circle, draw, fill=red!30] (F1) at (7,1) {4};

    \node[circle, draw, fill=green!30] (A1) at (5,2) {1};
    \node[circle, draw, fill=green!30] (B1) at (5,0) {2};
    \node[circle, draw, fill=green!30] (C1) at (6,1) {3};
    
    \node[circle, draw] (Ancestor) at (7,3.5) {};

    % Connect the nodes to the common ancestor with curved arrows
    \draw[<-, bend left] (A1) to (Ancestor);
    \draw[<-, bend left] (B1) to (Ancestor);
    \draw[<-, bend left] (C1) to (Ancestor);
    \draw[<-, bend right] (D1) to (Ancestor);
    \draw[<-, bend right] (E1) to (Ancestor);
    \draw[<-, bend right] (F1) to (Ancestor);

    \draw[color=gray!50,line width=0.1mm] (A1) -- (B1);
    \draw[color=gray!50,line width=0.1mm] (B1) -- (C1);
    \draw[color=gray!50,line width=0.1mm] (C1) -- (A1);
    \draw[color=gray!50,line width=0.1mm] (D1) -- (E1);
    \draw[color=gray!50,line width=0.1mm] (E1) -- (F1);
    \draw[color=gray!50,line width=0.1mm] (F1) -- (D1);
    \draw[color=gray!50,line width=0.1mm] (F1) -- (C1);

    \node at (6.1, -2) {\includegraphics[width=5cm]{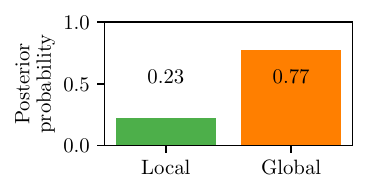}};

    \node at (3.5, -4) {$P(\blankcircle{1},\blankcircle{2},\blankcircle{3},\blankcircle{4},\blankcircle{5},\blankcircle{6})=\begin{cases}\frac{1}{Z}e^{J\sum_{ij} w_{ij} x_ix_j} \text{ (local)}\\\frac{1}{Z}e^{B\sum_{i}x_i} \text{ (global)}
     \end{cases}$};
\end{tikzpicture}
\caption{\textbf{Left}. Local coordination: nodes align to their neighbors through pairwise interactions. They may get stuck in a Nash equilibrium. %Locality implies that node 1 can only influence node 5 if the information propagates throughout intermediate nodes on a path between them (i.e. nodes 3 and 4).
\textbf{Right}. Global coordination: nodes are coordinated by a common cause transcending the graph structure (with some possible noise).
Local and global processes generally predict different patterns of coordination, which means their contribution can be inferred from behavioral data. In each of these toy examples of local and global coordination, the Ising model correctly identifies the most likely process. %Global mechanisms (driven by shared culture, common knowledge, or institutions) can enhance coordination in circumstances where it would be hard to achieve via local processes alone.
}
    \label{fig:local_global}
\end{figure}

Figure \ref{fig:local_global} illustrates how local and global processes may generate different patterns of coordination. %In an evolutionary game theoretic framework, locality implies that agents may only update their strategy based on the behavior of their own neighbors.
In this particular example, local coordination fails to produce consensus as the network is stuck into a suboptimal Nash equilibrium. Occasionally, ``global'' processes may solve this type of failure. Alternatively, local and global forces may push in opposite directions and complicate the emergence of a norm \citep{lee2006reconsideration} -- for instance, if different groups with incompatible inclinations come into contact. Figure \ref{fig:local_global} also shows that the Ising model can accurately infer the actual coordination process for each toy example.

In \S\ref{section:local_versus_global}, using an Ising model, we measure the contribution of local ($J$) and global ($\bm{B}$) mechanisms to the formation of physicists' preferences. We find evidence for both local and global effects in the case of the metric signature, while the latter seem to pre-dominate. Moreover, as will be shown in \S\ref{section:inferring}, this Ising model approach can serve as a basis for comparing the plausibility of more realistic mechanisms of preference-formation, according to whether they generate local or global coordination patterns. %In particular, we consider three different mechanisms for the formation of scientists' preferences (without making a strong commitment to any of them), and assess their relative plausibility according to their ability to account for the measured values of $J$ (local coordination) and $\bm{B}$ (global coordination). The first mechanism is an agent-based model that assumes physicists operate a trade-off between coordination costs, switching costs, and maladaptation costs. The second mechanism is a process of global cultural transmission (capturing, for instance, cultural transmission via textbooks) and the third mechanism considers local cultural transmission (via the imitation of peers), a channel that has the potential to propagate conventions \citep{Moore2012}.

\subsection{\label{section:optimality_costs}Optimality versus decision costs in the resolution of conflicts}

In the absence of norms, how can individuals with conflicting preferences eventually achieve coordination? In scientific collaborations, authors must sometimes overcome such conflicts. This can involve a trade-off between ``optimality'' (e.g. the maximization of their collective satisfaction), and ``decision costs''. Indeed, co-authors can seek to maximize their collective satisfaction through proper aggregation of their individual preferences. However, this can be cumbersome: not all decisions deserve to be put under the whole collective's scrutiny, and it might be easier to let a leader decide, potentially at the expense of collective agreement. It is indeed well known that power and leadership can mitigate decision and coordination costs \citep{williamson1975markets,calvert1992leadership}. In \S\ref{section:conflict_resolution}, we infer the mechanisms via which physicists resolve conflicting preferences towards the metric signature in co-authored papers. We find some evidence that leadership also plays a role in the resolution of conflicts, resulting in suboptimal decisions given that individual preferences are only partially aggregated within collaborations. While dictatorial strategies of conflict-resolution fail to represent individual preferences, they may occasionally produce more mutually coherent collections of decisions than e.g. majority voting \citep[p.~23]{Dietrich2006}.

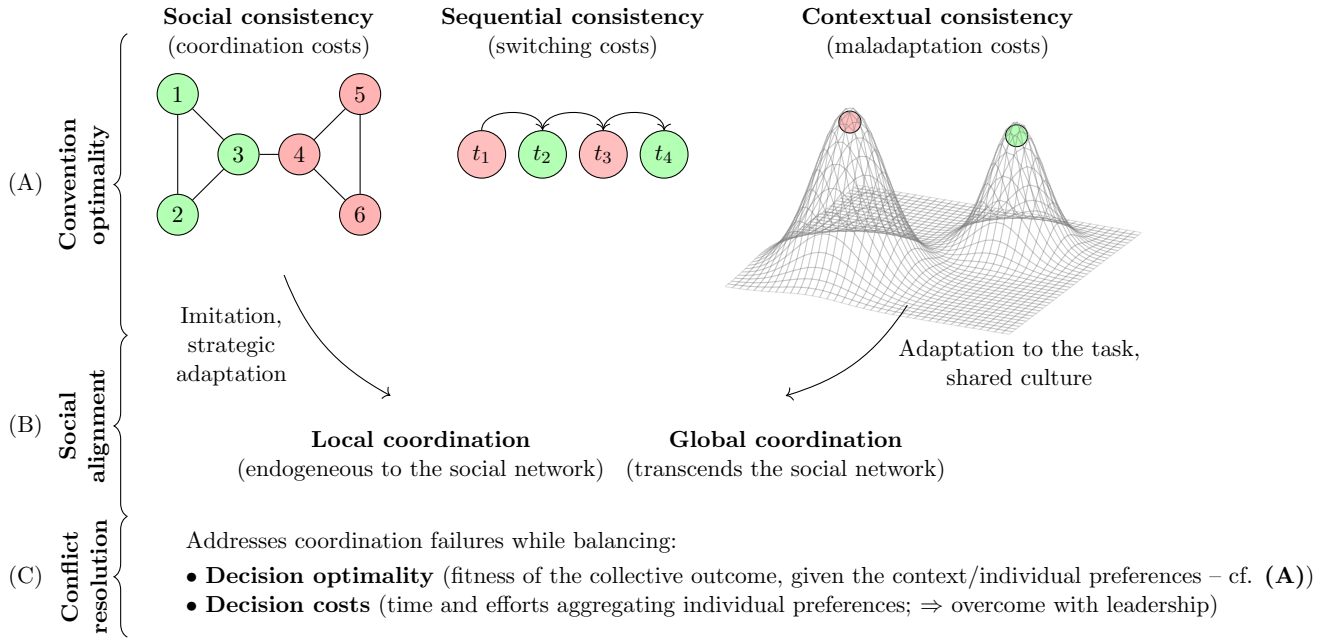
\begin{figure}[!h]
    \centering
\begin{tikzpicture}[scale=0.8, transform shape]

\draw [decorate,decoration={brace,amplitude=5pt,raise=4ex}]
  (-1,1) -- (-1,6) node[midway,rotate=90,yshift=+4em,align=center]{\textbf{Convention}\\\textbf{optimality}};

\begin{scope}[shift={(0,3)}]
    \node[circle, draw, fill=red!30] (D) at (2,2) {5};
    \node[circle, draw, fill=red!30] (E) at (2,0) {6};
    \node[circle, draw, fill=red!30] (F) at (1,1) {4};

    % Left graph (traditional)
    \node[circle, draw, fill=green!30] (A) at (-1,2) {1};
    \node[circle, draw, fill=green!30] (B) at (-1,0) {2};
    \node[circle, draw, fill=green!30] (C) at (0,1) {3};

    \draw (A) -- (B);
    \draw (B) -- (C);
    \draw (C) -- (A);
    \draw (D) -- (E);
    \draw (E) -- (F);
    \draw (F) -- (D);

    \draw (F) -- (C);

    \node[align=center] (label) at (0.5,3) {\textbf{Social consistency}\\(coordination costs)};
\end{scope}

\begin{scope}[shift={(4,4)}]
    \node[draw,circle,fill=pink] (a) at (0,0) {$t_1$};
    \node[draw,circle,fill=green!30] (b) at (1,0) {$t_2$};
    \node[draw,circle,fill=pink] (c) at (2,0) {$t_3$};
    \node[draw,circle,fill=green!30] (d) at (3,0) {$t_4$};
    \draw[->,bend right=90] (a.north) to [out=90,in=90] (b.north);
    \draw[->,bend right=90] (b.north) to [out=90,in=90] (c.north);
    \draw[->,bend right=90] (c.north) to [out=90,in=90] (d.north);

    \node[align=center] (label) at (1.5,2) {\textbf{Sequential consistency}\\(switching costs)};
\end{scope}

\begin{scope}[shift={(8,0)}]
\begin{scope}[shift={(0,0.75)}]
\begin{axis}[hide axis]
\node[draw,circle,fill=pink] at (axis cs:-1, -1, +1.05) {};
\node[draw,circle,fill=green!30] at (axis cs:1, 1, +0.80) {};
\addplot3 [
domain=-2.5:2.5,
domain y = -2.5:2.5,
samples = 40,
samples y = 40,
surf,
opacity=0.25,
fill=white,
fill opacity=0.01,
faceted color = gray] {+1.15*exp(-((x+1)^2+(y+1)^2)/0.67)+0.9*exp(-((x-1)^2+(y-1)^2)/0.4)};
\end{axis}
\end{scope}
\node[align=center] (label) at (3.5,6) {\textbf{Contextual consistency}\\(maladaptation costs)};
\end{scope}

\draw [decorate,decoration={brace,amplitude=5pt,raise=4ex}]
  (-1,-2) -- (-1,1) node[midway,rotate=90,yshift=+4em,align=center]{\textbf{Social}\\\textbf{alignment}};

\begin{scope}[shift={(3,-1)}]
    \node[align=center] (local) at (0,0) {\textbf{Local coordination}
    \\(endogeneous to the social network)}; 
\end{scope}

\begin{scope}[shift={(9,-1)}]
    \node[align=center] (global) at (0,0) {\textbf{Global coordination}\\(transcends the social network)}; 
\end{scope}

\draw[->, bend right=20] (0.75, 2) to node[align=center,xshift=-4em] {Imitation,\\
strategic\\adaptation} (2.5,0);

% \draw[<->, bend left=20] (5.25, 2.75) to (3.5,0);
% \draw[<->, bend right=20] (5.75, 2.75) to (8,0);
% \node[rotate=-90] at (5.5, 1) {Disrupt};

\draw[->, bend left=20] (11, 1.5) to node[align=center,xshift=+7em] {Adaptation to the context,\\
shared culture} (9,0);

\begin{scope}[shift={(-1,-3)}]
    \node[align=left,anchor=west] at (0,1.5em) {Addresses coordination failures while balancing:};
    
    \node[align=left,anchor=west] (optimality)    at (0,0) {$\bullet$ \textbf{Decision optimality} (fitness of the collective outcome, cf. \textbf{(A)})};
    
    \node[align=left,anchor=west] (costs) at (0,-1.25em)  {$\bullet$ \textbf{Decision costs} (time and efforts aggregating individual preferences)};
    
\end{scope}

\draw [decorate,decoration={brace,amplitude=5pt,raise=4ex}]
  (-1,-4) -- (-1,-2) node[midway,rotate=90,yshift=+4em,align=center]{\textbf{Conflict}\\\textbf{resolution}};

\node at (-3.5, 3.5) {(A)};
\node at (-3.5, -0.5) {(B)};
\node at (-3.5, -3) {(C)};

\end{tikzpicture}
    \caption{Three trade-offs (A,B,C) affecting conventions and their relationships.}
    \label{fig:trade-offs}
\end{figure}

\

We have identified three trade-offs affecting conventions. Figure \ref{fig:trade-offs} summarises these trade-offs and highlights their interactions. In what follow, we show how our statistical physics framework can be deployed to provide empirical evidence for these trade-offs.

\section{\label{section:results}An empirical case-study}

We turn to the empirical case-study of the metric signature convention in physics. To this end, literature in high-energy physics was collected from the Inspire HEP database, which includes various metadata (authorship, institutional affiliations, etc.). When available, the LaTeX source of each paper is retrieved from arXiv. 22500 papers from four categories (Phenomenology-HEP, Theory-HEP, General Relativity \& Quantum Cosmology, and Astrophysics) are successfully classified into either metric signature ($\pm 1$) using a small set of regular expressions (see S\ref{regular_expressions}).

\subsection{\label{section:consistency_context}Beyond coordination: the role of sequential and contextual consistency}

We have postulated that in addition to social coordination, individuals' attitude towards conventions may also be influenced by imperatives of sequential and contextual consistency. If sequential consistency matters, individuals should tend to use the same convention throughout their own works. By contrast, if individuals behave differently across research areas, we may infer that they value contextual consistency. %A crude measure of consistency is the auto-correlation $\langle x_{i,t} x_{i,t+1} \rangle = 0.72$.

\begin{figure}[!h]
    \centering
    \begin{subfigure}[b]{0.48\textwidth}
        \centering
        \includegraphics[width=0.85\textwidth]{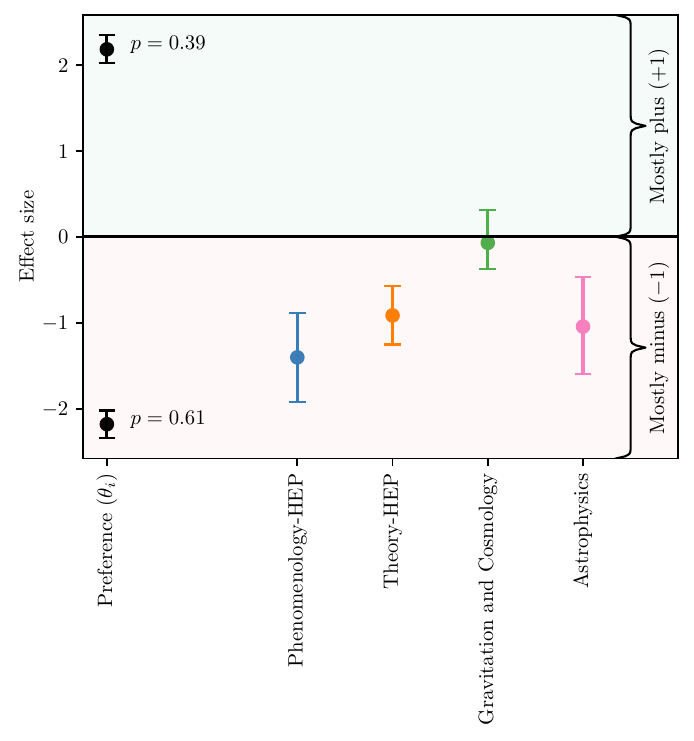}
        \caption{\label{fig:consistency_vs_task}Effect of sequential consistency (i.e. preferences, in black), and context (in color), on the choice of a convention in solo-authored papers. $p$ indicates the prevalence of each preference ($\pm 1$).}
    \end{subfigure}\hfill\begin{subfigure}[b]{0.48\textwidth}
        \centering
        \includegraphics[width=0.85\textwidth]{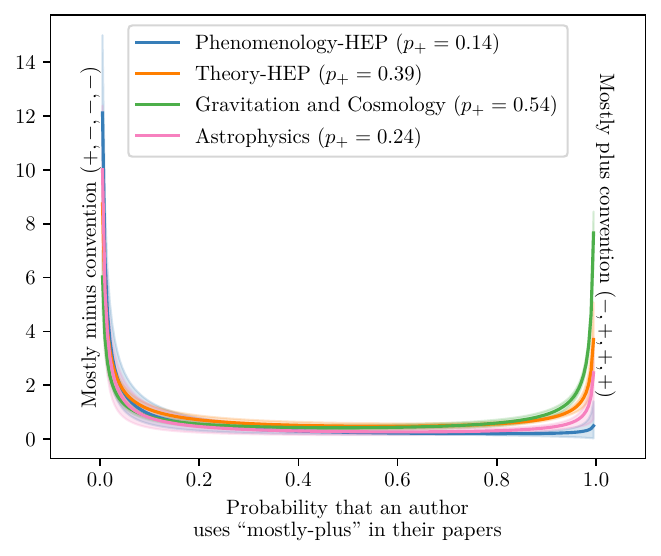}
        \caption{\label{fig:preferences}Distribution of authors' probability of using the mostly plus metric signature, according to their primary subject category. Shaded areas represent 95\% credible intervals (CI). Distributions are generally bimodal, with two peaks at 0 and 1, which imply that authors tend to use always one or the other signature but rarely a mix of both in solo-authored papers.}
    \end{subfigure}
    \caption{\textbf{Importance of sequential and contextual consistency in scientists' behavior.}}
\end{figure}

Below, we measure the importance of sequential and contextual consistency in scientists' behavior. We consider only solo-authored papers, for which the choice of metric purely reflects the sole author's choice. In order to capture the imperatives of sequential and contextual consistency, we assume that the probability that an author 
$i$ uses the $+1$ sign convention in a paper $d$ is:

\begin{equation}
    P(x_d=+1|i,c) = \frac{e^{\frac{1}{2}(\theta_i + b_c)}}{e^{\frac{1}{2}(\theta_i + b_c)}+e^{-\frac{1}{2}(\theta_i + b_c)}}% = \text{logit}^{-1} (\theta_i + b_c)
\end{equation}

where $\theta_i$ is a latent parameter that encodes author $i$'s preference ($\theta_i>0$ implying a preference for the $+1$ convention) and $b_c$ is a latent parameter that encodes the bias associated with context $c$ (the category of literature to which the paper belongs\footnote{In case a paper belongs to multiple categories, we average $b_c$ over all these categories.}). In our account of conventions, $\theta_i$ is the mean-field effect of sequential consistency, and $b_c$ is the mean-field effect of cultural traits interacting with the choice of the metric signature, given their distribution in a research area $c$. We assume that $\theta_i$ is drawn from a mixture of two distributions ($\theta_i=\pm \mu$), such that the model may capture the existence of two populations with a preference for each metric. We also assume that $b_c\sim\mathcal{N}(0,1)$\footnote{We assume that:
\vspace{-1em}
\begin{equation*}
\theta_{i} =
    \begin{cases}
      +\mu & \text{with probability } p_{C_i}\\
      -\mu & \text{with probability } 1-p_{C_i}
    \end{cases}
\end{equation*}
\vspace{-1em}
\\
where $C_i$ is the primary research area of author $i$ and $\mu\sim \text{Exponential}{(1)}$. The ability of this item-response model to reconstruct the latent parameters $\mu$ and $b$ is tested with simulated data assuming no effect of sequential consistency, i.e. $\theta_i=0$ for every author (S\ref{consistency_context_latent}, Figure \ref{fig:preferences_simulations}).
}. If $|\mu|$ is typically large, and larger than $|b|$, this would imply that scientists have preferences that generally exceed the influence of the context. As shown in Figure \ref{fig:consistency_vs_task}, we find that scientists \textit{do} have preferences that they tend to maintain across contexts, although there is some evidence that they occasionally adapt to the target research area. While we interpret such deviations from an author's preference as adaptation to the subject matter, they could indicate adaptation to the audience of the paper, in pursuit of social consistency (code-switching). %

Figure \ref{fig:preferences} \footnote{Given $N_i$, the amount of solo-authored papers by an author $i$ with an explicit choice of metric signature, and $k_i$ the amount of those using the $+1$ convention, we assume that $k_i \sim \text{Binomial}\left(N_i, p_i\right)$, with $p_i \sim \text{Beta}\left( \alpha_{C_i},\beta_{C_i}\right)$ and $\alpha_c,\beta_c \sim \text{Exponential}(1)$. %If $\alpha_c,\beta_c<1$, then authors from a research area $c$ tend to stick to their favorite metric.
} confirms that authors tend to generally stick to the same metric in their works and that the prevalence of each preference varies depending on the authors' primary research area. This shows that authors manage the tension between sequential and contextual consistency by developing preferences adapted to their cultural environment.

\subsection{\label{section:local_versus_global}Local versus global coordination: an Ising model approach}

\begin{figure}[p]
    \centering
    \includegraphics[width=0.9\linewidth,trim=250 350 225 350,clip]{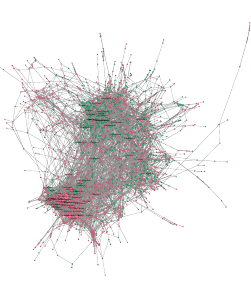}
    
    \caption{\textbf{Metric signature preferences in the co-author network}. Each node is an author. Edges represent co-authorship relationships between authors. Nodes' colors indicate authors' preferences (pink for $-1$, green for $+1$). Only the largest connected component is shown. 
    }
    \label{fig:co-author-network}
\end{figure}

If scientists' attitude towards the sign convention was dictated by social consistency, then, their preferences should be aligned with their social environment. While there exists no universal norm at the level of the entire field, it could still be the case that scientists are at least behaving in a way consistent with their own collaborators. To establish whether this is the case, we explore the co-authorship graph (Figure \ref{fig:co-author-network}), where each node $i$ on the graph (each author) possesses a favorite convention $x_i \in \{\pm 1\}$  (as measured from their solo-authored publications). The weights of the edges ($w_{ij}$) encode the strength of the relationship between co-authors $i$ and $j$\footnote{We use $w_{ij} = \sum_{d \mid \{i,j\} \subset A_d} \frac{1}{|A_d|-1}$, where $A_d$ is the set of co-authors of publication $d$, following \citep{Newman2004}.}. We may then measure the average alignment between co-authors, $\langle x_i x_j \rangle = \sum_{i,j} w_{ij} x_i x_j/\sum_{i,j} w_{ij}$. We find $\langle x_i x_j \rangle = +0.32$, which is significantly more than would be expected by chance alone ($P<10^{-4}$)\footnote{We compare the observed value of $\langle x_i x_j \rangle$ to what would be expected if authors chose one or the other convention at random, with probabilities equal to the frequency of each convention. This null model predicts $\mathbb{E}[\langle x_i x_j\rangle]=0.10$, far below the observed value. %The non-zero prediction of the null model stems from the fact that the mostly minus convention is more common, such that co-authors'  preferences would be somewhat aligned even by chance alone.
}: despite the absence of universal norm, scientists' preferences are positively correlated with those of their collaborators.

How did such partial alignment emerge? Coordination among physicists may be achieved either locally (via short-range interactions between scientists), or globally, via shared culture. 
% The largest connected component of the co-authorship graph exhibits two dense clusters (Figure \ref{fig:co-author-network}), one dominated by authors with a preference for the mostly minus signature (in pink), and one dominated by the opposite signature (in green). It is unclear, however, whether such agreement within clusters is explained by local interactions, or by a cultural predisposition that is widely shared \textit{within} each cluster but differs across clusters.
To delineate these two possibilities, we model physicists' preferences with an Ising model, with parameters $J$ and $\bm{B}$, such that the probability $P(x_1,\dots,x_n|J,\bm{B})$ of observing a particular configuration $x_1,\dots,x_n$ is:

\begin{align}
    \label{eq:ising}
    P(x_1,\dots,x_n|J,\bm{B}) &= \frac{1}{Z(J,\bm{B})} e^{U(x_1,\dots,x_n,J,\bm{B})}\\ \text{ with } U &= \underbrace{\sum_{i,j} J w_{ij} x_{i}x_j}_{\substack{\text{local}\\\text{coordination}}} + \underbrace{\sum_i B_{C_i}x_i}_{\substack{\text{global}\\\text{coordination}}}
\end{align}

Where $C_i$ is the primary research area of $i$. $J$ captures the effect of \textit{local} coordination via pairwise interactions on the graph. $\bm{B}=(B_c)$ captures the \textit{global} effect of each research area: their effect is global in that they equally affect all individuals within a group regardless of their position in the network. %Using the cultural landscape analogy previously evoked, $\bm{B}$ can be interpreted as the ``mean-field'' effect of other cultural traits associated with the preference for a convention over the other, given their distribution in a particular research area.
%Consequently, each author experiences two influences: that of their social environment (via $J$), and that of their broader research area (via $\bm{B}$). In the Ising model, $U$ is the potential of a particular configuration, as defined by the values of $x_i$ for every $i$.
The Ising model follows naturally from eq. \eqref{eq:coordination}, \S\ref{paragraph:coordination} in coordination games. The $\bm{B}$ term introduces an asymmetry between authors from different research areas\footnote{Unlike Table \ref{table:coordination_game}, we assume that the effect of the asymmetry between research areas does not scale linearly with each node's degree centrality ($k_i=\sum_j w_{ij}$). Instead, each strategy is associated with a constant payoff $r_i=B_{C_i}x_i$ regardless of the interactions involving $i$ \citep{Zimmaro2024}}.

If $J>0$, the potential $U$ is higher in configurations in which nodes share the orientation of their neighbors. Such systems may undergo phase transitions towards configurations in which individual nodes spontaneously align over large distances.
Although originated from spin physics, the Ising model provides a concise description of the emergence of collective behavior at large \citep{Macy2024,Galesic2019}.

\begin{figure}[!h]
    \centering
\resizebox{0.65\linewidth}{!}{
\begin{tikzpicture}

    % Parallelogram for Layer 1 (very transparent gray with more skew)
    \filldraw[fill=gray!10, draw=none] 
    (-2,-0.5) -- (3,-0.5) -- (4,2.5) -- (-1,2.5) -- cycle;

    % Parallelogram for Layer 2 (very transparent gray with more skew)
    \filldraw[fill=gray!10, draw=none] 
    (-2,3) -- (3,3) -- (4,6) -- (-1,6) -- cycle;

    % Left graph (traditional)
    \node[circle, draw, fill=red!30] (A) at (0,0) {4};
    \node[circle, draw, fill=green!30] (B) at (2,0) {5};
    \node[circle, draw, fill=red!30] (C) at (1,2) {2};
    \node[circle, draw, fill=green!30] (D) at (3,2) {3};
    \node[circle, draw, fill=red!30] (F) at (-0.5,1) {1};

    \node (g) at (-2,1) {\Large $G$};

    \node[circle, draw, fill=red!30] (A1) at (0,3.5) {4};
    \node[circle, draw, fill=green!30] (B1) at (2,3.5) {5};
    \node[circle, draw, fill=red!30] (C1) at (1,5.5) {2};
    \node[circle, draw, fill=green!30] (D1) at (3,5.5) {3};
    \node[circle, draw, fill=red!30] (F1) at (-0.5,4.5) {1};

    \node (gcit) at (-2,4.5) {\Large $G^{\text{cit}}$};

    \draw (A) -- (B);
    \draw (C) -- (D);
    \draw (C) -- (F);

    \draw[color=gray!70,line width=0.5mm,dashed] (A) -- (A1);
    \draw[color=gray!70,line width=0.5mm,dashed] (B) -- (B1);
    \draw[color=gray!70,line width=0.5mm,dashed] (C) -- (C1);
    \draw[color=gray!70,line width=0.5mm,dashed] (D) -- (D1);
    \draw[color=gray!70,line width=0.5mm,dashed] (F) -- (F1);

    \draw[->] (C1) -- (A1);
    \draw[->] (F1) -- (A1);
    \draw[->] (D1) -- (B1);

    \node at (7, 2) {\includegraphics[width=5.5cm]{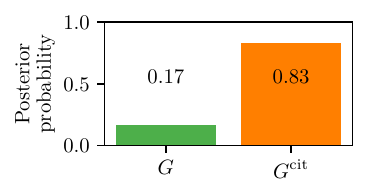}};

    \node at (3.5, -1.75) {$P(\blankcircle{1},\blankcircle{2},\blankcircle{3},\blankcircle{4},\blankcircle{5})=\begin{cases}\frac{1}{Z}e^{J\sum_{ij} w_{ij} x_ix_j} 
 \ (G)\\\frac{1}{Z}e^{J\sum_{ij} w_{ij}^{\text{cit}} x_ix_j} 
 \ (G^{\text{cit}})\end{cases}$};
    
\end{tikzpicture}
}
\caption{\textbf{Illustration of local coordination in multilayered social networks}. Nodes can be connected through different kinds of relationships (for instance, authors can be related via collaborations ($G$) or citations ($G^{\text{cit}}$)). In this diagram, patterns of coordination are better explained by the directed graph at the top ($G^{\text{cit}}$). The Ising model correctly identifies the relevant social structure.}
    \label{fig:multilayered}
\end{figure}

%In our case, we would like to infer the posterior distribution $P(J,B|x_1,\dots,x_n)$ given $(x_1,\dots,x_n)$. However, this distribution is computationally intractable, and we use the pseudo-likelihood approximation \citep{Nguyen2017} which is accurate, efficient, and robust to missing data as we show in S\ref{ising_robustness}. The results are shown in Table \ref{table:ising}. The inverse Ising approach reveals that research areas have large global effects, and that local coordination of co-authors has a small but statistically significant effect.
We infer the posterior distribution $P(J,B|x_1,\dots,x_n)$ given $(x_1,\dots,x_n)$\footnote{See \citep{Nguyen2017} and S\ref{ising_robustness}.}. The results are shown in Table \ref{table:ising}. The inverse Ising approach reveals that research areas have large global effects, and that local coordination of co-authors has a small but statistically significant effect. However, this convention may propagate locally via channels others than collaborations, including citations (Figure \ref{fig:multilayered}). We account for this possibility by introducing an additional local contribution $J^{\mathrm{cit}} \sum_{j} w^{\text{cit}}_{ij}x_j$ in the pseudo-likelihood (\eqref{eq:pseudo_likelihood}), induced by the authors' citation graph $G^{\text{cit}}$ which captures ``who cites who''\footnote{The pseudo-likelihood approach can directly accommodate asymmetric interactions in directed networks.}. The weights $w^{\text{cit}}_{ij}$ of the edges of $G^{\text{cit}}$ measure the frequency of citations of $j$ by $i$, given $w^{\text{cit}}_{ij}=\sum_{d,d'|i\in A_d, j\in A_{d'},i\neq j} \frac{c_{dd'}}{|A_d||A_{d'}|}$ with $c_{dd'}=1$ if $d$ cites $d'$ and 0 otherwise. After adding this contribution to \eqref{eq:pseudo_likelihood}, we find that both $J$ and $J^{\mathrm{cit}}$ are significantly positive; that is, both co-authors and publications seem to carry an influence\footnote{That $J$ remains positive after accounting for citations suggests that correlations between co-authors' preferences may not be explained solely by correlations in their research.}. 

To assess which of local or global coordination dominate, we evaluate the fraction of authors for which local contributions in \eqref{eq:pseudo_likelihood} exceed the global effect of $\bm{B}$. We find that local effects exceed and reverse global effects for 7\% of the sample of $2\,277$ authors (CI$_{\text{95\%}}=[$3\%--15\%]). In addition, we find that the inclusion of local effects only marginally improves the model's predictive accuracy, from an average of 67.7\% (only considering global effects) to 70.2\%. Therefore, local processes play a smaller role.

Measurements of $J$ and $\bm{B}$ may be confounded by hidden structures. For instance, while $\bm{B}$ was assumed to be uniform within each of the four research areas, it may vary across subtopics within each research areas. If their effect is omitted, this might inflate the estimate of $J$; upon further analysis, this possibility is excluded\footnote{To assess this possibility, a linear Support Vector Machine classifier was trained to predict the metric signature from sentence embeddings of scientific abstracts. The accuracy on a test-set was 73\%, slightly above the accuracy of a classifier relying only on the four categories (70\%); in other words, categories contain most of the relevant contextual information.}. Conversely, the effect of each research area may reflect unmodeled social structures. Therefore, the Ising model is an effective parameterization, and the values of $J$ and $\bm{B}$ may vary depending on the networks and scales under consideration. Finally, when the social structure is entirely correlated with another interacting trait, their relative effect cannot be teased apart. 
%Fortunately, this approach is flexible: it can incorporate any combination of networks and simultaneously infer their relative contribution (Figure \ref{fig:multilayered}).

%For instance, we might want to add a ``co-journal network'', encoding how often a pair of physicists have published in identical journals. We may also consider a co-institution graph, encoding whether scientists have been simultaneously employed at the same institution\footnote{The result will also depend on the weights of each graph's edges, and one could argue for many different choices. In the general formulation of the inverse Ising problem, the underlying graph edges weights are also to be inferred from the data. This requires, however, the observation of many different configurations. In our case, by definition, only one configuration is observed.}.
%Similarly,\footnote{If the partition is too-fined grained, then $\bm{B}$ has too many dimensions for them to be effectively measured; however, the issue can be alleviated by a hierarchical model, assuming that each research area is divided into smaller topics and so forth, and that deviations of each child-topic around their parent-topic are small. }. 

\subsection{\label{section:inferring}Inferring mechanisms of preference formation}

The Ising model is certainly not a realistic description of the process through which individuals form preferences. Nevertheless, idealized models from statistical physics can provide clues about the actual process. To demonstrate how, we assess the relative plausibility of three hypothetical mechanisms according to their ability account for the observed magnitudes of local coordination ($J$) and global coordination ($\bm{B}$). %Although none of these may be compelling accounts of reality, they will primarily serve to illustrate how to achieve some understanding of the underlying mechanism of preference-formation given the balance between local and global coordination.

The first proposed mechanism ($M_1$) is an agent-based model in which scientists operate a trade-off between social consistency (driven by coordination costs), sequential consistency (driven by switching costs), and contextual consistency (driven by maladaptation costs, i.e. incompatibility with their research area). In this model, the network is initialized in a random state; then, at every step of the simulation, scientists follow a best response strategy, by evaluating whether they would be better off changing their preference or not, given the magnitude of each of these costs, their probability of publishing in each research area, and their collaborators' preferences\footnote{See S\ref{section:strategic_agent_model} for a more precise description.} (in that scenario, coordination is channeled by co-authorship and not citations). The second mechanism considered ($M_2$) is a global process of cultural transmission whereby scientists adopt a convention at the start of their career with a probability that depends on their primary research area, and on the time at which their career started. Such process is meant to capture the transmission of conventions via cultural artefacts such as textbooks (S\ref{global_transmission}). Finally, the third mechanism considered ($M_3$) is a process of local cultural transmission, in which scientists copy the preference of their first co-author\footnote{The preference of scientists with no ``parent'' in the graph is drawn according to the same global process as in the global cultural transmission model ($M_2$), such that the process $M_3$ includes both local and global mechanisms. In total, in this model, 10\% of authors form a preference by imitation.}. 

% Many samples are drawn according to each generative process $M_1,M_2,M_3$. For each sample, we infer the parameters of the Ising model ($\bm{B}$, $J$ and $J^{\mathrm{cit}}$) -- ignoring the authors whose actual preference is unknown, in order to preserve the compatibility with the values of $\bm{B}$, $J$ and $J^{\mathrm{cit}}$ inferred from the actual data). Since each model generates slightly different patterns for these parameters (Figure \ref{fig:all}), these can be used as summary statistics for estimating their relative plausibility given the observed data, $P(M|J,J^{\mathrm{cit}},\bm{B})$. For this task, we use simulation-based inference \citep{Cranmer2020} with BayesFlow \citep{radev2021amortized,radev2023bayesflow}. This procedure allows to perform Bayesian inference when one lacks an analytical expression for the likelihood $P(D|M)$, and all that can be done is drawing samples by simulating the generative process $M$. This technique is especially useful for making inferences about models defined by complex programs, such as agent-based models% \citep{Shiono2021}
% . When the data is highly dimensional (as in the present case), this approach requires ``summary statistics'' \citep{Cranmer2020}. Interestingly, the parameters of the Ising model can serve this role. Figure \ref{fig:confusion_matrix} confirms that the procedure exhibits some ability to discriminate the three models.

\begin{figure}[h!]
    \centering
    \includegraphics[width=0.99\linewidth]{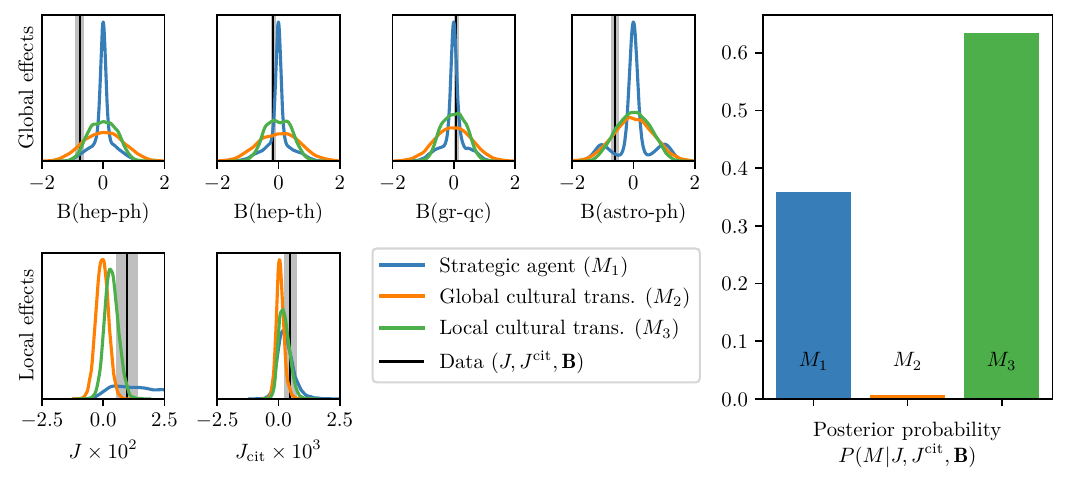}
    \caption{\textbf{Left plot:} marginal posterior distribution of summary statistics for each model (shown in colors), compared to the summary statistics derived from the data (indicated by black bars). Gray bars represent the 95\% posterior credible interval of each parameter given the data. \textbf{Right plot:} posterior probability of each model given the observed parameters of the Ising model.}
    \label{fig:all}
\end{figure}

Figure \ref{fig:all} shows that each model predicts different patterns for $J$ and $\bm{B}$. In particular, since it explicitly implements coordination costs (which are themselves driven by local interactions), the ``strategic agent'' model can predict large values of $J$. The model of cultural transmission via imitation predicts slightly higher values of $J$ than global cultural transmission, but generally smaller values of $B$. Because of these distinctive patterns, we can compare each model's ability to account for the data using simulation-based inference \citep{Cranmer2020}. As shown in Figure \ref{fig:all}, the results seem to rule out purely global cultural transmission which fails to explain the magnitude of local coordination. There is slightly more support for a local cultural transmission model.

\subsection{\label{section:conflict_resolution}Inferring mechanisms of conflict resolution}

Coordination failures give rise to conflicts. Given that physicists' preferences are not perfectly aligned to those of their collaborators, they must occasionally resolve disagreements about which metric signature to use as they co-author a paper. We stressed that the resolution of conflicts in such scenarios implied a trade-off between optimality and decision costs: while some decisions may be superior to others, the cost of arguing and properly aggregating each author's input may exceed the benefits. 

Below, we consider multiple preference aggregation strategies and estimate their prevalence given data about the metric signature selected in co-authored papers. As we will show, this provides indirect information about how authors navigate this trade-off in the case of the metric signature. We leverage papers with an identified metric signature $S\in\{\pm 1\}$ for which all authors' preferences $(x_i,\dots,x_n)\in\{\pm 1\}^n$ were measured independently from single-authored papers. For many of these papers (182 papers with two authors, 28 papers with three authors, and 4 papers with four authors), authors have conflicting preferences. Since different processes of preference-aggregation occasionally predict different outcomes given $(x_i,\dots,x_n)\in\{\pm 1\}^n$, we may infer their relative likelihood from the data.

In contrast to \citep{Rotabi2017}, which explored conventions for LaTex macros, we consider strategies of conflict-resolution suggested by the literature on judgment aggregation \citep{arrow1951social,list2011group}. First, we consider ``dictatorial'' strategies, whereby a specific author (the first author, the last author, or any other one) imposes their favorite convention (which, again, is independently measured from their solo-authored publications). Dictatorial strategies dismiss all information about other authors or the research context, such that the resulting decision is potentially suboptimal. We also consider a ``majoritarian'' process, whereby the majority preference is selected, thus maximizing collective satisfaction. These two strategies (dictatorial and majoritarian) are probably the most classic examples in social choice theory and in the preference and judgment aggregation literature \citep{arrow1951social,list2011group}. It is also tempting to consider the achievement of consensus through deliberation, another popular example. However, it seems difficult to infer whether a decision was reached from deliberation based solely on the observed outcome and each individual's initial preference. Instead, we consider a ``random'' process, equivalent to a coin-flip (in fact, in the two-author case, a coin-flip is presumably equivalent to deliberation, if both authors are equally influential in the deliberation). Finally, we include a ``conventional'' process, whereby the signature most frequent in a given context is retained, irrespective of the authors' preferences. We then estimate the prevalence prevalence $\pi_k$ of each preference aggregation strategy $A_k \in \{A_1,\dots\}$, given that $P(S|x_1,\dots,x_n) = \sum_k P(S|x_1,\dots,x_n,A_k) P(A_k)$, and $A_k \sim \mathrm{Categorical}(\pi_k)$.

\begin{figure}[h]
    \centering
    \includegraphics[width=0.8\linewidth]{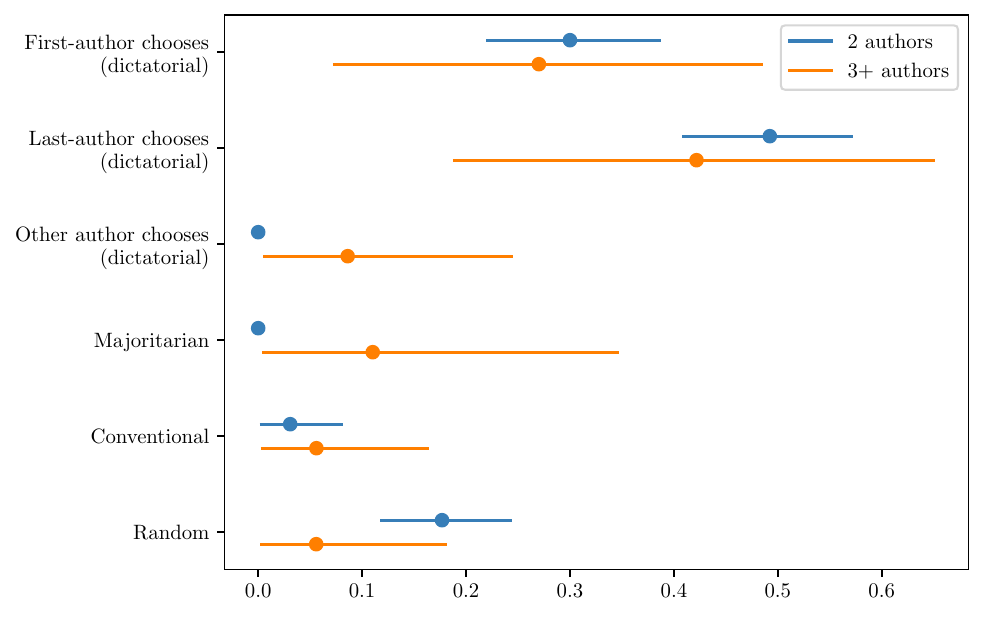}
    \caption{Prevalence of aggregation strategies. Error bars indicate 95\% credible intervals. The dominant strategy seems to be that the last author dictates the metric convention.}
    \label{fig:aggregation}
\end{figure}

Results are shown in Figure \ref{fig:aggregation}, given a flat Dirichlet prior on $\pi_k$. Due to the sample size, error bars are quite wide. Nevertheless, dictatorial strategies prevail ($\pi_{\text{dictatorial}}>0.73$ at the 95\% credible level for the two-author case and $\pi_{\text{dictatorial}}>0.57$ for the three+-author case -- which is almost always three authors), even in the $3+$ authors case (for which majority vote is possible): inequalities in the status of authors can help expedite judgment aggregation. More interestingly, in the two-author configuration, there is conclusive evidence that the first author is less likely to choose the metric signature compared to the last author ($P(\pi_{\text{first-author}}>\pi_{\text{last-author}})=0.008$). For $3+$ authors, the data leans towards this direction as well ($P(\pi_{\text{first-author}}>\pi_{\text{last-author}})=0.222$); moreover, middle-authors seem less likely to dictate the final choice. The last author (who is generally in a leadership position) therefore seems to enjoy more influence, even though the first author carries a greater share of work (in principle) and would benefit from using their favorite metric signature\footnote{Authorship norms are known to vary across fields \citep{waltman2012empirical}. To verify that these interpretation hold in fundamental physics, we evaluated the probabilities that the first-author or the last-author are strictly older than the other co-authors. We found an association between last-authorship and seniority (see S\ref{authorship_norms} for more details).}. This converges with \citep{Rotabi2017}, which finds that ``fights'' for visible conventions in scientific papers are more often won by experienced authors.
This emphasizes the role of leadership in the resolution of conflicts, and suggests that for this particular convention, ``optimality'' (whether in the sense of promoting collective agreement, or the first author's satisfaction) is sacrificed. 
%The exact process through which the last author decides is unclear %; they might be choosing the target journal, which might itself have a favorite metric signature\footnote{This hypothesis can in principle be tested by comparing the correlation between the target journal and each author's previous papers' journals.}.
%and probably implicit.
%Since the last author's preference effectively prevails more often, we may conclude that highly influential authors are protected from coordination costs and that conventions in co-authored papers are not representative of their authors' average preference. 
%Finally, the fact that power shields senior authors from coordination costs raises further questions such as: can conventions propagate when the most central nodes in a network (assuming centrality is associated with power) do not experience coordination costs? 

\section{\label{section:discussion}Conclusion}

This paper introduced a statistical physics account of conventions capturing the heterogeneous, competing processes influencing their adoption. This account acknowledges that individuals attitude' towards conventions is dictated not only by an imperative of social coordination, but also by purely internal imperatives of sequential consistency (the need to avoid switching back-and-forth between different choices) and contextual consistency (the need to adopt mutually coherent systems of choices, as in epistemological holism; \citealt{ben2006conventionalism}). Broadly speaking, conventionality arises when a collection of choices or traits is constrained only collectively, due to synergistic interactions between individual choices. Interestingly, all three imperatives (social, sequential, and contextual consistency) can be modeled, at least approximately, by coordination games on some underlying graph structure. We thus provide a mereological account of conventions, as arising from a specific type of interaction between parts, resulting in constraints not on the individual parts themselves but rather on the whole. We highlight a connection with the Ising model, established in the context of ``potential games'', that can be leveraged to extract information about the drivers underlying behavioral patterns in real-life conventions. As we showed, this can provide information about the structure of the underlying social coordination game and the social networks involved. Additionally, for systems of interacting conventions, the Ising model can reveal the underlying cultural fitness landscape.

Additionally, the framework implies that social coordination can arise out of either \textit{local} (or bottom-up) coordination, driven by dyadic interactions on a network, or from \textit{global} (top-down) coordination, relying on shared culture and knowledge or institutions transcending the social infrastructure. The contribution of these two processes in real-life conventions can also be measured with an Ising model approach. Additionally, different models of preference-formation (imitation, adaptation, shared cultural artifacts, etc.) predict different magnitudes of local and global coordination, such that the magnitude of these two channels of coordination can help determine the actual underlying process in real conventions. In the case of the metric signature, we found slightly more evidence in favor of cultural transmission of preferences via the imitation of a peer, a process that can explain a small but non-vanishing magnitude of local coordination. %Purely global cultural transmission (as one might expect from the imitation of textbooks) is ruled out due to its inability to account for the observed magnitude of local coordination. However, our work did not exhaust all possible mechanisms, which was not our aim. This would require more realistic models and additional summary statistics beyond the parameters of the Ising model. For instance, different processes of preference-formation might predict different levels of intra- and inter-generational coordination, and this could be leveraged in their comparison. Nevertheless, the local versus global distinction is generally insightful.
In scientific communities, it may explain which aspects of epistemic cultures belong to a ``disciplinary matrix'' \citep{kuhn1970} (the set of practices and values that scientists adopt as part of the process of acquiring a disciplinary identity) and which aspects emerge more spontaneously and locally. %More generally, we show how the Ising approach provides a relatively model-independent way of discriminating local (i.e. emergent and endogenous) from global (exogenous) collective synchronization using behavioral network data.

Finally, it was argued that the resolution of conflicts between multiple conventions implies a trade-off between the optimality of the outcome (e.g., the degree of collective satisfaction) and decision costs (i.e. the cost of reaching a decision). We illustrated this trade-off using the example of the metric signature. We inferred the prevalence of various preference-aggregation strategies in co-authored papers, and found more evidence for ``dictatorial'' strategies. Specifically, we found that the last-author's preference has a higher chance of prevailing, leading to suboptimal outcomes. Therefore, leadership and seniority play a role in addressing coordination problems in the absence of norm. 

Empirically, owing to its statistical framing, the framework provides an array of tools for understanding either the lack of norm or the persistence of inferior norms and practices in a wide range of contexts. In certain context, this may require adjustments. For instance, while the Ising model presupposes pairwise interactions (between, say, individuals, or cultural traits), complex systems often involve higher-order interactions. In particular, scientists frequently interact in collaborations involving more than three authors, for which they might develop specific conventions. Such complex interactions, whether they involve multiple individuals or cultural traits, can be encoded by hyper-graphs \citep{Zimmaro2024}.  In our framework, this leads to a generalized Ising model \citep{Robiglio2024}. The mereological perspective is useful in that respect, since it highlights the common structure underlying higher-order interactions across systems, suggesting modeling strategies that may equivalently represent game-theoretic interactions or cultural fitness landscapes \citep{Jansma2025}. Moreover, although this paper limits itself to a binary convention, the approach can be extended to conventions involving more than two alternatives, as in the Potts model \citep{wu1982potts}.

%Additionally, to our knowledge, this paper is the first attempt to reverse-engineer the processes of judgment-aggregation in co-authored publications \citep{Bright2017}, and our approach may be applied to many other decisions.
%Finally, in contrast to \citep{Rotabi2017}, this paper has not paid much attention to temporal dynamics, due to the temporal sparsity of the data. Nevertheless, exploring such dynamics would provide more information about the underlying processes of transmission, or about how sequential consistency plays out over time. 

%TC:ignore

% \paragraph{Acknowledgements} Many thanks to Radin Dardashti, Thomas Heinze, Olivier Morin, and Cailin O'Connor for their feedback. I am also grateful to Jeffrey Barrett, Sean Carroll, John Miller, Scott Page, and Brian Skyrms for inspiring discussions. Finally, I would like to thank Robert D. Hawkins for further suggestions.
\paragraph{Acknowledgements} [ANONYMIZED]

% \paragraph{Code and data} The code and data for this paper is available at \url{https://gin.g-node.org/lucasgautheron/dilemmas-conventions}.
\paragraph{Code and data} The code and data for this paper is available at [ANONYMIZED].

% \paragraph{Funding} The author acknowledges funding from the DFG Research Training Group 2696 ``Transformations of Science and Technology since 1800''.
\paragraph{Funding} [ANONYMIZED]

\paragraph{Competing interests} The author declares a personal inclination towards the $(+,-,-,-)$ metric signature.

\printbibliography

\newpage 

\appendix
\section{Supplementary materials}

\subsection{\label{regular_expressions}Regular expressions for determining the metric signature}

The following case-insensitive regular expressions have been used to detect occurences of the mostly minus signature:

\begin{itemize}
    \item \texttt{(([,\textbackslash{}s\textbackslash{}\{\textbackslash{}\}]*)(\textbackslash{}+|1)([,\textbackslash{}s\textbackslash{}\{\textbackslash{}\}1]*))\{1\}(([,\textbackslash{}s\textbackslash{}\{\textbackslash{}\}]*)\textbackslash{}-([,\textbackslash{}s\textbackslash{}\{\textbackslash{}\}1]*))\{3,\}
}
    \item \texttt{(mostly[-\textbackslash{}s]*minus|west[-\textbackslash{}s]*coast)
}
    \item \texttt{g\_\textbackslash{}\{(00|tt)\textbackslash{}\}[\textbackslash{}s]*=[\textbackslash{}s]*[+]?[\textbackslash{}s]*1
}
    \item \texttt{\textbackslash{}\textbackslash{}Box(\textbackslash{}\^{}(\textbackslash{}\{2\textbackslash{}\}|2))?[\textbackslash{}s]*\textbackslash{}+[\textbackslash{}s]*m\textbackslash{}\^{}(\textbackslash{}\{2\textbackslash{}\}|2)
}
\end{itemize}

Symmetric expressions are conversely employed for detecting instances of the mostly plus metric signature.

\subsection{\label{landscapes}Reconstructing cultural landscapes of conventions}

\begin{figure}[H]
    \centering
    \begin{subfigure}[b]{0.48\textwidth}
        \centering
        \includegraphics[width=0.95\textwidth]{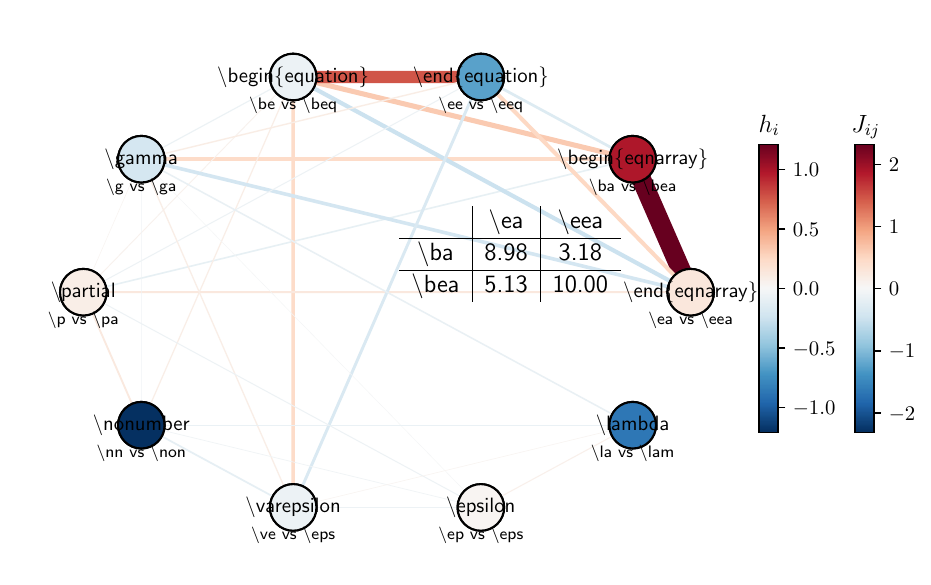}
        \caption{Ising model representation of a cultural landscape involving ten binary conventions, following \citep{Poulsen2023}. Each convention represents a choice between two options to shorten the name of a LaTeX command. Edges represent interactions between conventions. Thick edges designate traits that must be mutually consistent. Node colors indicate the intrinsic advantage of a specific choice (blue indicates that the shorter abbreviation is favored, red indicates a general preference for the longer abbreviation). Interactions between two traits can be seen as coordination games. The inverse Ising problem recovers the coefficients of the underlying payoff matrices.}
    \end{subfigure}\hfill\begin{subfigure}[b]{0.48\textwidth}
        \centering
        \includegraphics[width=1.1\textwidth]{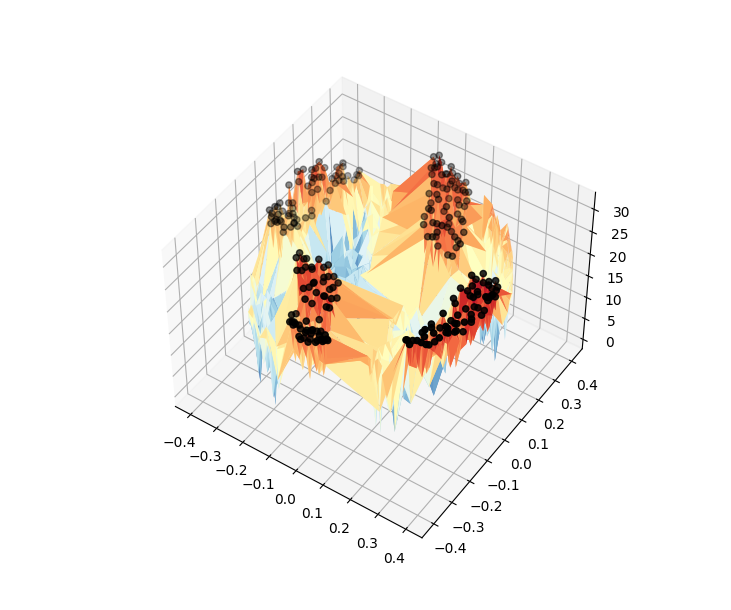}
        \caption{\label{fig:preferences}Three-dimensional representation of the cultural landscape. Each of the $2^{10}=1024$ potential combination of traits is mapped onto two dimensions using multidimensional scaling. The height and color of the landscape indicates the fitness of each combination. There are four large peaks (in red), corresponding to the most stable systems. The black dots represent the configurations in which the abbreviations for the pairs (\texttt{\textbackslash begin\{equation\}}, \texttt{\textbackslash end\{equation\}}) \& (\texttt{\textbackslash begin\{eqnarray\}},  \texttt{\textbackslash end\{eqnarray\}}) are consistent.} 
    \end{subfigure}
    \caption{\label{fig:landscapes}\textbf{Cultural landscape of common abbreviations of LaTeX commands.}.}
\end{figure}

This section illustrates the ability of the Ising model to reconstruct cultural fitness landscapes of conventions. To this end, following \citep{Rotabi2017}, we explore ten naming conventions involving LaTeX macros abbreviating the name of frequently used LaTeX commands. For instance, instead of writing \texttt{\textbackslash begin\{equation\}} at the beginning of each equation, authors often use an abbreviated name (e.g. \texttt{\textbackslash be}) by defining a custom macro (e.g. \texttt{\textbackslash newcommand\{\textbackslash be\}\{\textbackslash begin\{equation\}\}}). The choice of abbreviated name is conventional and in principle at the author's discretion. For instance, certain authors prefer ``\texttt{\textbackslash beq}'' over \texttt{\textbackslash be}. We collect occurrences of the two most frequent abbreviated names for each of ten such conventions, labeled by $-1$ (for the shorter version, for instance \texttt{\textbackslash be}) and $+1$ (for the longer version, for instance \texttt{\textbackslash beq}). Following our framework, we assume that the fitness of a combination of choices is given by the Ising model, $U(x_1,\dots,x_{10})=\sum_{ij} J_{ij}x_ix_j + \sum_i h_i x_i$. Assuming $J_{ij},h_i\sim\mathcal{N}(0,1)$ and, using data from 77\,000 papers in which at least one convention appears, we solve an inverse Ising problem to recover $(J_{ij})$ and $(h_i)$. The results are shown in Figure \ref{fig:landscapes}. This reveals strong interactions between two pairs of choices: the abbreviations of (\texttt{\textbackslash begin\{equation\}}, \texttt{\textbackslash end\{equation\}}), and those of (\texttt{\textbackslash begin\{eqnarray\}}, \texttt{\textbackslash end\{eqnarray\}}).

% \subsection{\label{replication}Replication analysis: the choice between python and R}

% \begin{figure}[H]
%     \centering
%     \begin{subfigure}[b]{0.48\textwidth}
%         \centering
%         \includegraphics[width=0.95\textwidth]{replication/consistency_vs_task_simple.pdf}
%         \caption{\label{fig:consistency_vs_task}Effect of sequential consistency (i.e. preferences, in black), and context (in color), on the choice of a convention in solo-authored papers. $p$ indicates the prevalence of each preference (python versus R).}
%     \end{subfigure}\hfill\begin{subfigure}[b]{0.48\textwidth}
%         \centering
%         \includegraphics[width=0.95\textwidth]{replication/preferences.pdf}
%         \caption{\label{fig:preferences}Distribution of authors' probability of using python over R, according to their primary subject category. Shaded areas represent 95\% credible intervals (CI). Distributions are generally bimodal, with two peaks at 0 and 1, which imply that authors tend to use always either python or R but rarely a mix of both in solo-authored papers.}
%     \end{subfigure}
%     \caption{\textbf{Importance of sequential and contextual consistency in scientists' choice between python and R.}}
% \end{figure}

\subsection{\label{consistency_context_latent}Sequential versus contextual consistency: model assessment}

% \begin{figure}[H]
%     \centering
%     \begin{subfigure}[T]{0.48\textwidth}
%         \centering
%         \includegraphics[width=0.95\textwidth]{consistency_vs_task.pdf}
%         \caption{\textbf{Main plot}: effect of individual preferences and context on convention choice. The brown curve represents the distribution of individual preferences ($\theta<0$ indicates a preference for $-1$, and $\theta>0$ for $+1$). Vertical bars represent $b_c$, the effect of each research area (context). Shaded areas indicate 95\% CIs. Individual preferences dominate (on average, $|\theta|>|b|$).
%         \textbf{Top-right plot}: comparison of $b_c$ across the four research areas. Errors bars indicate 95\% CIs. There is some evidence of an effect of context in phenomenology and gravitation \& cosmology.}
%         \label{fig:preferences}
%     \end{subfigure}\quad\begin{subfigure}[T]{0.48\textwidth}
%         \centering
%         \includegraphics[width=0.95\textwidth]{consistency_vs_task_simulation.pdf}
%         \caption{The same analysis is re-iterated with simulated data instead of actual data. The simulation assumes that $\theta_i=0$ for all authors (i.e. there is no effect of consistency), while each research area has a significant effect. The inference correctly finds that $|\theta|$ is typically lower than $|b|$ and correctly identifies the true size of the effect of each research area.}
%         \label{fig:preferences_simulations}
%     \end{subfigure}
%     \caption{\textbf{Importance of consistency and context in scientists' choice of convention.}}
% \end{figure}

\begin{figure}[H]
    \centering
    \includegraphics[width=0.45\textwidth]{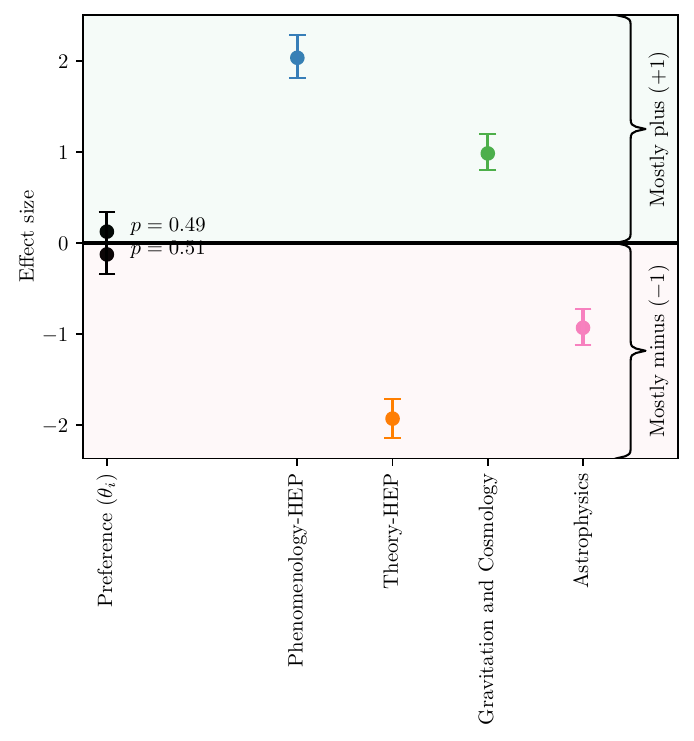}
    \caption{\label{fig:preferences_simulations}
The analysis in \ref{section:consistency_context} is re-iterated with simulated data instead of actual data. The simulation assumes that $\theta_i=0$ for all authors (i.e. there is no effect of consistency), while each research area has a significant effect. The inference correctly finds that $|\theta|$ is nearly zero and correctly identifies the ground truth size of the effect of each research area (+2, -2, +1, and -1 respectively).}
\end{figure}

% \subsection{\label{ising_justification}}

\subsection{\label{ising_robustness}Inverse ising problem and the pseudo-likelihood approach}

The pseudo-likelihood method \citep{Nguyen2017} transforms the inverse Ising problem into a tractable logistic regression, based on the likelihood of observing each individual spin conditional on the others, i.e.:

\begin{equation}
    \label{eq:pseudo_likelihood}
    \prod_{i} P(x_i=+1|\{x_{j\neq i}\}) = \prod_{i} \dfrac{e^{+J\sum_j w_{ij} x_j + B_{C_i}}}{e^{+J\sum_j w_{ij} x_j + B_{C_i}}+e^{-J\sum_j w_{ij} x_j - B_{C_i}}}
\end{equation}

\begin{figure}[H]
    \centering
    \includegraphics[width=0.8\linewidth]{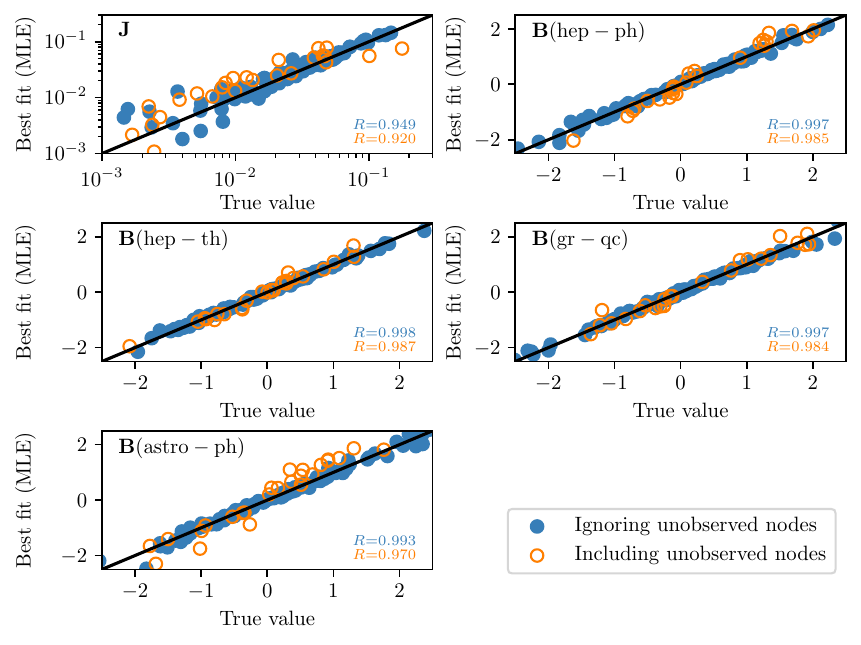}
    \caption{Robustness of the pseudo-likelihood approach for measuring $J$ and $\bm{B}$. ``True'' values of $J$ and $\bm{B}$ are drawn at random [$J\sim\text{Exponential}(1/J^{\ast})$, $\bm{B}\sim\mathcal{N}(0,1)$]. Node configurations $(x_i)$ are drawn at random  according to the Ising model for each values of $J$ and $\bm{B}$, using Gibbs sampling, either i) removing or ii) including nodes corresponding to authors whose preference is not observed in the data. Finally, the maximum likelihood estimates (MLE) $J^{\text{MLE}}$ and $\bm{B}^{\text{MLE}}$ are recovered with the pseudo-likelihood approach, for each configuration $(x_i)$, imputing $x_i=0$ for authors whose preference was not observed in our data. The best-fit values are in reasonably good agreement with the true values over the simulated range, although they are much less accurate in the case where unobserved authors are included in the Gibbs sampling process.}
    \label{fig:ising_robustness}
\end{figure}

Using simulated configurations of $G$, we demonstrate that the pseudo-likelihood approach provides reliable estimates of $J$ and $\bm{B}$, if all $x_j$ are observed, and for $J\leq 10^{-2}$ (Figure \ref{fig:ising_robustness}). In the case that a value $x_j$ is unknown, due to a lack of paper solo-authored by $j$ with an identified metric signature, then author $j$ is omitted from the sums in \eqref{eq:pseudo_likelihood}. This is equivalent to imputing $x_j = 0$\footnote{This imputation strategy is also equivalent to restricting the inference procedure to a sub-graph of the co-authorship graph, including only the nodes and edges involving the $2\,277$ authors whose preference could be identified in at least one solo-authored paper.}. We find that this approach is able to recover reliable information about the true value of $J$ (Appendix \ref{ising_robustness}, Figure \ref{fig:ising_robustness}). However, we may fear that the imputation of missing data (equivalently interpretable as the removal of unobserved nodes from the network) introduces bias in our inference \citep{Poulsen2023}. A proper handling of unknown authors' preferences would require marginalizing eq.~\eqref{eq:pseudo_likelihood} over the $2^{m}$ possible combinations of the $m$ underlying unobserved signatures\footnote{An alternative would be Gibbs sampling, which may handle missing data without marginalization, though it turned out to perform worse than HMC in the present case.}. Unfortunately, the amount of missing data makes this impossible. However, this issue is not necessarily critical if, ultimately, we are less interested in recovering the exact values of $J$ and $\bm{B}$ than in using the estimates as summary statistics for the purpose of comparing multiple models of the formation of individual preferences. Then, as long as each model predicts distinct patterns for the best-fit values of $J$ and $\bm{B}$, the procedure remains useful. In any case, simulations show that the measured value of $J$ is very correlated with the true value, even when nodes with missing data are masked during the inference process (cf. Appendix \ref{ising_robustness}, Figure \ref{fig:ising_robustness}). Finally, missing data could be a feature rather than a bug; they might manifest the fact that certain authors make no explicit use of a specific metric signature, in which case it is reasonable to assume that they may not exert any influence over their co-authors' preferences.

\begin{table}[H]
\centering
\begin{tabular}{lcccc}
\toprule
 & Effect size & CI$_{\text{95\%}}$ & Effect size & CI$_{\text{95\%}}$ \\
Parameter &  &  &  &  \\
\midrule
$J$ & +0.013 & [+0.009, \ +0.017] & +0.0095 & [+0.0052, \ +0.014] \\
$J^{\mathrm{cit}}$ & - & - & +0.00049 & [+0.00023, \ +0.00075] \\
$B(\mathrm{hep-ph})$ & -0.86 & [-0.99, \ -0.73] & -0.77 & [-0.91, \ -0.64] \\
$B(\mathrm{hep-th})$ & -0.22 & [-0.29, \ -0.15] & -0.17 & [-0.24, \ -0.095] \\
$B(\mathrm{gr-qc})$ & +0.075 & [-0.0069, \ +0.16] & +0.076 & [-0.0066, \ +0.16] \\
$B(\mathrm{astro})$ & -0.6 & [-0.74, \ -0.47] & -0.59 & [-0.73, \ -0.46] \\
\bottomrule
\end{tabular}
\caption{\label{table:ising}Parameters of the Ising model.}
\end{table}

\subsection{\label{preference_formation_model}Models of preference formation}

Three models are considered: a strategic agent model ($M_1$), a global-transmission model ($M_2$), and a local transmission model ($M_3$). Many samples are drawn according to each generative process $M_1,M_2,M_3$. For each sample, we infer the parameters of the Ising model ($\bm{B}$, $J$ and $J^{\mathrm{cit}}$) -- ignoring the authors whose actual preference is unknown, in order to preserve the compatibility with the values of $\bm{B}$, $J$ and $J^{\mathrm{cit}}$ inferred from the actual data). Since each model generates slightly different patterns for these parameters (Figure \ref{fig:all}), these can be used as summary statistics for estimating their relative plausibility given the observed data, $P(M|J,J^{\mathrm{cit}},\bm{B})$. For this task, we use simulation-based inference \citep{Cranmer2020} with BayesFlow \citep{radev2021amortized,radev2023bayesflow}. This procedure allows to perform Bayesian inference when one lacks an analytical expression for the likelihood $P(D|M)$, and all that can be done is drawing samples by simulating the generative process $M$. This technique is especially useful for making inferences about models defined by complex programs, such as agent-based models% \citep{Shiono2021}
. When the data is highly dimensional (as in the present case), this approach requires ``summary statistics'' \citep{Cranmer2020}. Interestingly, the parameters of the Ising model can serve this role. Figure \ref{fig:confusion_matrix} confirms that the procedure exhibits some ability to discriminate the three models.

\subsection{\label{section:strategic_agent_model}Strategic agent model}

The ``strategic agent'' model proceeds as follow:

\begin{enumerate}
    \item The parameters of the model are drawn at random:
    \begin{itemize}
        \item $c_{b} \sim \mathcal{N}(0,1)$, defined for each research area $b$, is the (dis)advantage of the $+1$ convention in $b$. The cost of using a convention $x$ in context $b$ is $\max(0, -x c_{b})$.
        \item $c_c \sim \text{Exponential}(\langle d_i \rangle)$ represent the magnitude of coordination costs, where $\langle d_i \rangle$ is the average degree-centrality of authors in the co-authorship graph. The mean is thus set such that $\langle c_c\rangle \langle d_i \rangle=1$.
        \item The cost of switching from one convention to another is fixed ($c_s=1$)\footnote{This breaks a degeneracy of the model due to scale-invariance (if all costs were rescaled by a certain quantity, agents' behavior would remain identical).}.
    \end{itemize}
    \item At $t=0$, the network is initialized in a random state: $x_{i,t=0}$ is set to either $-1$ or $+1$ with equal probabilities.
    \item At $t+1$, each agent compares their payoff in two scenarios: i) they switch their preference ($x_{i,t+1}=-x_i$) or ii) they maintain it ($x_{i,t+1}=x_i$). The difference in payoffs is:
    {
    \small
    \begin{equation}
        \Delta = -c_s - c_c \sum_j w_{ij} \left(\max(0,x_{j,t} x_{i,t})-\max(0,-x_{j,t} x_{i,t}) \right) - \sum_{b} p_{ib} \left(\max(0, x_{i,t} c_{b})-\max(0, -x_{i,t+1} c_{b})\right)
    \end{equation}
    }
    
    Where $p_{ib}$ is the probability that $i$ publishes in research area $b$. If $\Delta>0$, $i$ switches their preference. The cost of switching ($c_s$) introduces an asymmetry in $\Delta$ and has the effect of a conservative bias.
    
    \item The process is repeated 50 times. The amount of steps reflects a compromise between performance and convergence.
    
\end{enumerate}

This best-response strategy model is similar to common logit-response approaches to belief dynamics such as \citealt{Galesic2021}, in the limit $\beta\to+\infty$ (see eq. 1.6).

\subsection{\label{global_transmission}Global transmission model}

For the global transmission model, we assumed that the probability of adopting a specific convention depends on both time and the author's primary research area. The time-dependence was captured by a random walk. The rate of change in the random walk was obtained by fitting the model to data on reference books for which approximate patterns of citations throughout time could be measured. We manually determined the metric convention used in each of these references. These gave us a measure of the prevalence of each convention in the citations of reference textbooks' throughout time. Unfortunately, this measure itself was too imperfect to reflect the actual probability that a scientist adopts a convention from a specific textbooks. Nevertheless, we used the rate of variation of this measure with time in our random walk model.

\subsection{\label{sbi_pairplot}Distribution of summary statistics across models}

\begin{figure}[H]
    \centering
    \includegraphics[width=0.5\linewidth]{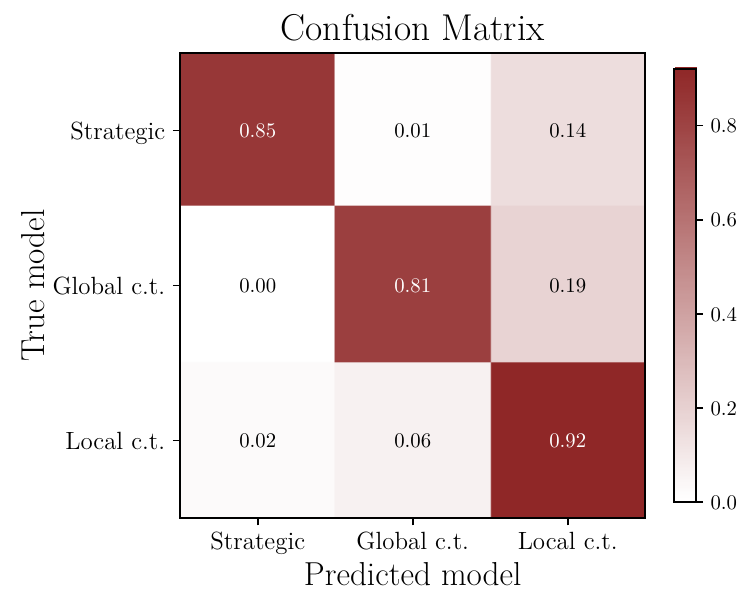}
    \caption{Reliability of the simulation-based model comparison approach. The confusion matrix represents the probability that a sample drawn from the horizontal model is attributed to the vertical model.}
    \label{fig:confusion_matrix}
\end{figure}

Conditioning the outcome of simulations on high-dimensional data $D$ to evaluate $P(\cdot|D)$ is difficult because the probability of generating exactly $D$ becomes virtually zero. One should therefore condition on summary statistics $T$ living in a lower dimensional space. Ideally, the mapping $f:D\mapsto T$ should be chosen in a way that maximizes our ability to tell apart the hypotheses that we seek to discriminate. In our case, $f:(x_1,\dots,x_n)\mapsto J,\bm{B}$ may not be optimal in that specific sense, but it has \textit{some} discriminating power (see Figure \ref{fig:confusion_matrix}) and has the merit of interpretability. A trivially better summary statistic for assessing the plausibility of, say, the model of local cultural transmission would be, for instance, the average rate of agreement between each author's and their first co-author (whose preference they should have imitated, according to the model). 

\begin{figure}[H]
    \centering
    \includegraphics[width=0.9\linewidth]{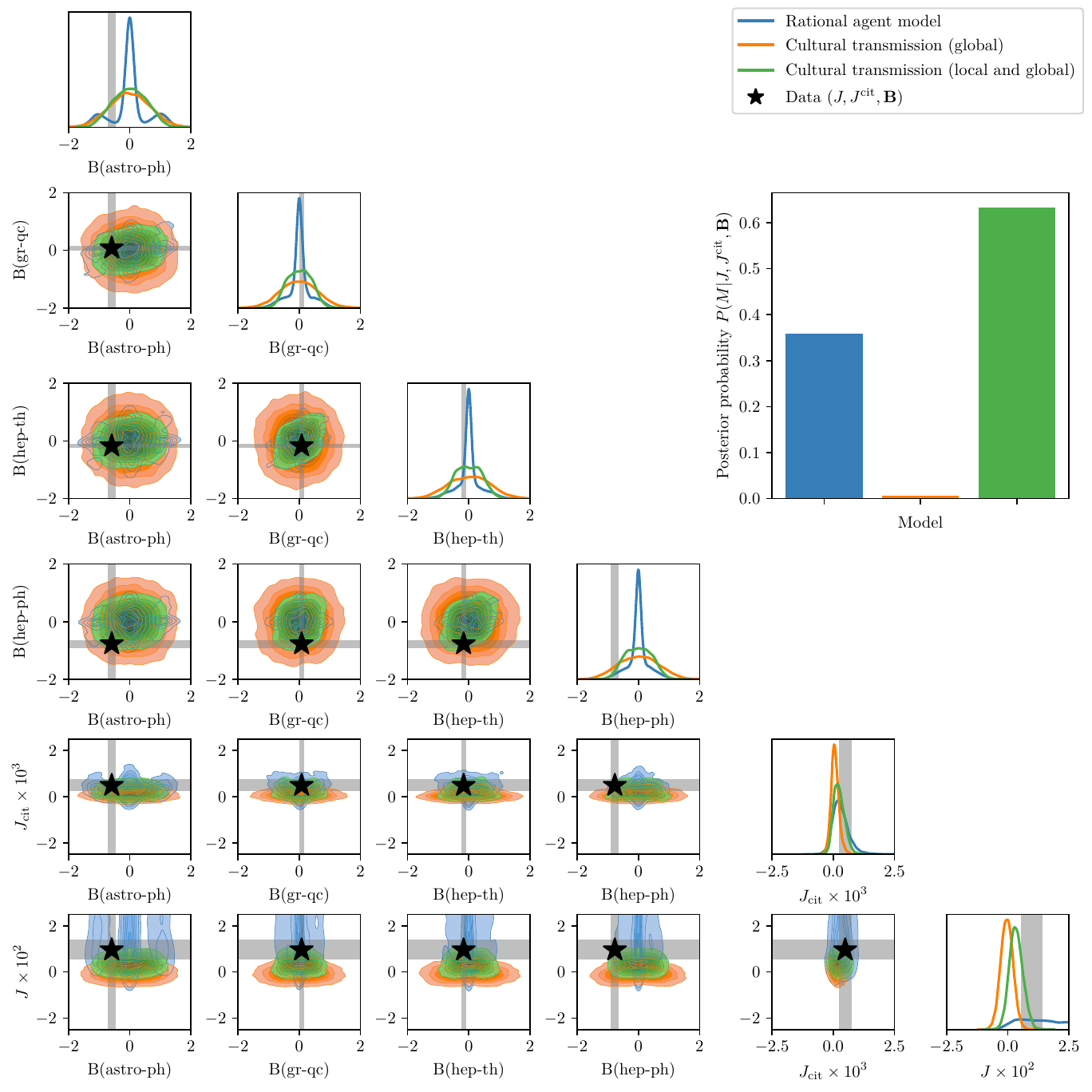}
    \caption{\textbf{Bottom-left pair plot:} distribution of summary statistics for each model (shown in colors), compared to the summary statistics derived from the data (shown as black stars). Plots on the diagonal show the marginal posterior distribution of each summary statistics for each model (gray bars represent the 95\% posterior credible interval of each parameter given the data). \textbf{Top-right bar plot:} posterior probability of each model given the observed parameters of the Ising model.}
    \label{fig:all_pairplot}
\end{figure}

\subsection{\label{authorship_norms}Authorship norms}

We investigated authorship norms in fundamental physics (excluding experimental physics, which are not considered in this paper and have very unusual norms). We found that the author-list of 79\% of two-author papers are alphabetically ordered. Given that for $n$ authors, there is a $1/(n!)$ chance that any ordering is equal to the alphabetical order, this implies that 56\% of two-author papers author-lists are \textit{intentionally} ordered \citep{waltman2012empirical}. This number goes down to 45\% for four-author publications. Therefore, despite a high prevalence of alphabetical ordering in fundamental physics compared to other disciplines (as found by \citealt{waltman2012empirical}), in about half of the publications the ordering of authors is meaningful.

Most importantly, we found evidence that last-authorship is associated with seniority: in 54\% of two-author papers, the last author has an academic age strictly higher than the first author; in comparison, in only 40\% of cases, the first-author has strictly higher seniority compared to the last-author.  In the three-author case, the last author has the strictly highest seniority in 29\% of cases, versus 17\% for the first-author.

%TC:endignore

\end{document}